\documentclass[preprint,11pt]{elsarticle}

\usepackage[left=3cm,right=3cm]{geometry}

\usepackage{graphicx}
\usepackage[caption=false, subrefformat=parens,labelformat=parens]{subfig}
\usepackage[percent]{overpic}

\usepackage{amsmath}

\usepackage{multirow}
\usepackage{color}

\usepackage{appendix}

\usepackage[normalem]{ulem}
\usepackage{url}
\usepackage{booktabs}

\usepackage{hyperref}

\begin{document}

\begin{frontmatter}

\title{A New Traders' Game? ---\\ Empirical Analysis of Response Functions in a Historical Perspective}

\author{Cedric Schuhmann}\ead{cedricschuhmann@gmail.com}
\author{Benjamin Köhler}\ead{benjamin.koehler@uni-due.de}
\author{Anton J. Heckens}\ead{anton.heckens@uni-due.de}
\author{Thomas Guhr}\ead{thomas.guhr@uni-due.de}
\address{Fakult\"at f\"ur Physik, Universit\"at Duisburg--Essen, Duisburg, Germany}

\begin{abstract}
Traders on financial markets generate non--Markovian effects in
various ways, particularly through their competition with one another
which can be interpreted as a game between different (types of)
traders. To quantify the market mechanisms, we empirically analyze self--response
functions for pairs of different stocks and the corresponding trade
sign correlators. While the non--Markovian dynamics in the
self--responses is liquidity--driven, it is expectation--driven in the
cross--responses which is related to the emergence of correlations. We empirically study the non--stationarity of these responses over time. In our
previous data analysis, we only investigated the crisis year 2008. We now
considerably extend this by also analyzing the years 2007, 2014 and
2021. To improve statistics, we also work out averaged response
functions for the different years. We find significant variations over
time revealing changes in the traders' game.
\end{abstract}

\end{frontmatter}

\section{Introduction}

Understanding price formation is a key issue in economics and
finance~\cite{BOUCHAUD_2009,Bouchaud2023,Lillo2023}. Prices are
influenced by news or information on trades, orders and cancellations
that is available to all traders in the order
book~\cite{CutlerPoterbaSummers1989,Fair2002,Joulin2008}.  The main
dynamics of the price formation takes place at the best buy or sell
orders.

In the research field of
market microstructure~\cite{Hasbrouck_book_2007, OHara_2015, Krishnan_2024}, one of
the first models of how trading mechanisms influence
price formation
was put forward by Kyle~\cite{Kyle_1985}. In his foundational paper, the
author models the interplay of stylized types of traders, namely noise
traders,
a trader with insider information and a market maker. From this he
deduces that the price impact is scaling linear with the order flow.
Glosten and Milgrom model the influence of trades on the price formation
from the viewpoint of market makers, who want to protect themselves from
better informed traders \cite{GlostenMilgrom_1985}. Their work
connected the economic terms of adverse selection and information
asymmetry with the context of market microstructure. Some more recent
modeling efforts can be found in Refs.~\cite{ContDeLarrard_2013,
BacryMuzy_2014}. In the former, the authors use an analytically
tractable Markov queueing process to model a stylized order flow. The
latter models price changes and order flow with a multivariate Hawkes
process, to capture the self--exciting nature of financial markets.

There are two closely related empirical observations for individual
assets that are essential to understand price formation. First, there
are long--term memory effects in the order flow known as trade sign
self--correlations or autocorrelations in the order
flow~\cite{Bouchaud_2003,LilloFarmer+2004,Lillo_2005,gerig2008theorymarketimpactorder,TOTH_2015,BOUCHAUD_2018}. They
can be traced back to splitting of metaorders, also referred to as
hidden orders. Second, there are self--responses that quantify the
accumulated temporal influence of an incoming order and the subsequent
price change of individual assets. In Refs.~\cite{Chordia_2002,
Chordia_2004} the authors show empirical evidence for both, strong
autocorrelations between daily imbalances of buy and sell orders, as
well as correlations between returns and lagged order imbalance on a
horizon of one trading day. In Ref.~\cite{ContKukanovStoikov_2013} the
authors introduce the order flow imbalance, an empirical variable which
entails changes in bid and ask queues, and provide robust evidence of a 
linear relation between this variable and returns on a high--frequency
timescale.

These empirical findings can also be interpreted as a competition of
two types of traders, liquidity takers and liquidity providers
\cite{BOUCHAUD_2009,Bouchaud2023,Lillo2023,Bouchaud_2003}.
Liquidity takers split larger orders to not attract the attention by
other traders with the intention to keep price levels around the
midpoint constant. Their actions cause time--lagged correlations and
thus trade sign self--correlations. Liquidity providers try to
anticipate the liquidity takers' trading behavior and take advantage
of differences in the bid--ask spread by placing their orders at more
expensive price levels. Liquidity takers are often willing to pay for
the additional risk compensation. Due to non--stationarity in the
traders` liquidity game the shape of self--responses are concave or
increase for larger time lags~\cite{Bouchaud_2006}. However, the
intraday price returns show a diffusive behavior if a certain balance
between sub-- and superdiffusive dynamics exists for many stocks
because traders exploit (statistical) arbitrage opportunities.
This was analyzed in Ref.~\cite{WeberRosenow_2005}, where the
authors introduce the virtual price impact, the impact a trade would
have solely due to the current state of the order book. They show that
the volume imbalance of limit buy vs sell orders is negatively
correlated to the returns. This explains the difference between real and
virtual price impact, as incoming limit orders cancel the predictable
change in price due to trades. Thus market efficiency is violated on
shorter intraday time scales and restored on longer
ones~\cite{Bouchaud_2003,Fama1970}.

Similar empirical observations are also found when looking at
different asset pairs for the same asset class. In the economics
literature various kinds of cross--effects were identified in
Refs.~\cite{Chordia_2000, HASBROUCK_2001,
  Pasquariello_2013,
  Boulatov_2012}. In the
econophysics literature, empirical studies focus on the
cross--response and the trade sign cross--correlations, respectively.
Self-- and cross--responses and trade sign correlators together
dynamically reveal the non--Markovian effects that the market has on
itself~\cite{WSG2015preprint,Wang2016,Wang2016_2,Benzaquen_2017,WangGuhr_2017,Grimm2019,HenaoLondono2021,HENAOLONDONO_2022}.
These quantities are also closely related to the concept of
cross--impact
~\cite{Lillo2023,Benzaquen_2017,HASBROUCK_2001,SchneiderLillo_2019,Cont03102023,Coz_2024}.
The cross--impact contains information on a significant fraction of
return covariances~\cite{Benzaquen_2017}. Cross-response, trade sign
cross--correlations and the cross--impact reflect the emergence of
correlations which arise due to different traders' market expectations
and are thus expectation or information driven. The self--responses,
trade sign self--correlations and self--impact are liquidity,
\textit{i.e.} supply and demand,
driven~\cite{WSG2015preprint,Wang2016,Wang2016_2,Benzaquen_2017,WangGuhr_2017,Grimm2019,HenaoLondono2021,HENAOLONDONO_2022}.
Correlations in the order flows of different stocks are caused by
multi--asset trading such as portfolio rebalancing strategies and or
pair trading~\cite{capponi2020multi}. The cross--impact is not
directly needed for modeling correlations between returns and order
flows of different stocks, but the cross--responses also contain
important information on the traders' actions, and thus on the
non--stationarity in the trading. Further details have been explored,
such as the market collective response~\cite{WNG2018} and the
role of asymmetric information in market impacts~\cite{WNG2019}

Here, we extend the analysis of
Refs.~\cite{WSG2015preprint,Wang2016,Wang2016_2} for the year 2008 by
studying the years 2007, 2014 and 2021. In doing so, we
rely on the well--established and theoretically explored framework of
self-- and cross--response functions in returns and trade signs \cite{WSG2015preprint, Wang2016, Wang2016_2, Benzaquen_2017, WangGuhr_2017, HenaoLondono2021, HENAOLONDONO_2022}, which we apply in a large scale empirical analysis across the several distinct time periods. The comparison of the years mentioned above covers pre--crisis, crisis and post--crisis periods and also includes the rise of algorithmic trading. We empirically analyze and discuss our results regarding both, the market, as well as the sectoral level, and distinguish between intra--sectoral and inter--sectoral measures. There
were two intriguing observations for the crisis year 2008. First, the
whole market restored partially its efficiency in the course of a
trading day,
\textit{i.e.} the cross--response for the market reached a maximum
after some minutes, followed by a slowed
decrease~\cite{WSG2015preprint,Wang2016}. Second, the memory in the
order flow is long--term~\cite{Wang2016_2}. Here we show that the
global crisis of 2008 had a major and lasting effect on self-- and
cross--responses. Our goal is to provide an in depth empirical account using established empirical measures of market microstructure to study this non--stationarity of the
trading mechanism over different periods and to answer the
question whether or not the traders' game changed.

After briefly introducing our data set in Sec.~\ref{sec:DataSet}, we
define self-- and cross--responses and trade sign self-- and
cross--correlators in Sec.~\ref{sec:TimeScalTimeLCrossCorr}. The
results of our data analysis are presented in Sec.~\ref{sec:Results}.
Theoretical and Modeling Aspects are briefly discussed in Sec.~\ref{sec:TMA}.
We conclude with Sec.~\ref{Sec:Discussion} by discussing in depth
empirical evidence for a different traders' behavior.

\section{\label{sec:DataSet}Data set}

We use the Daily TAQ (Trade and Quote) of the New York stock exchange (NYSE)~\cite{NYSE_2014} for the years 2007, 2008, 2014 and 2021. This data set records the trading activity of most stocks on the US stock markets, \textit{i.e.} the prices are listed in the trade file and best bid and ask prices in the quote file.
The time period used for each trading day is set at 9:40 to 15:50
(local time of the stock market) in order to reduce opening, closing and overnight effects.

We also analyze the stocks as in Refs.~\cite{WSG2015preprint,Wang2016,Wang2016_2} from the NASDAQ stock exchange for 2008, \textit{i.e.} ten stocks with high liquidity from each of the ten economic sectors according to the Global Industry Classification Standard (GICS)~see Tab.~\ref{tab:GICS}. 
\begin{table}[htbp]
	\centering
	\caption{\label{tab:GICS}
		Global Industry Classification Standard (GICS).
	}
	\begin{tabular}{l}
		\toprule
		Industrials   \\
		Health Care  \\
		Consumer Discretionary   \\
		Information Technology  \\
		Utilities  \\
		Financials  \\
		Materials   \\
		Energy   \\
		Consumer Staples  \\
		Telecommunication Services  \\
		\bottomrule
	\end{tabular}
\end{table}
In the case of the telecommunications sector, the number of available stocks is limited to nine.
As criterion for liquidity the trading value, which is the price multiplied by the corresponding volume~\cite{DUARTEQUEIROS2016}, was used in Ref.~\cite{WSG2015preprint,Wang2016,Wang2016_2}.
To compare different years, our focus is on selecting the same 99 stocks as in 2008 for all years.
However, stocks may be delisted from the stock market due to liquidations, bankruptcy, mergers or acquisitions for the other three years.
We replace the stocks delisted in a respective economic sector by high--liquid stocks from the same economic sector, \textit{i.e} by stocks with a high trading value.
Compared to 2008, the years of 2007, 2014 and 2021 still have 85, 89 and 71 stocks in common, respectively.
For 2007, we selected eleven stocks from the materials sector and nine stocks from the health care sector.
For all four years, we group the stocks according to the GICS, see the details in~App.~\ref{sec:Tables}.

\section{\label{sec:TimeScalTimeLCrossCorr}Response Functions and Trade Sign Correlators}

We work on the physical time scale in discrete time steps of one second. 
For 2007 and 2008, one second is the finest available time grid resolution possible~\cite{WSG2015preprint,Wang2016,Wang2016_2}.
For the years 2014 and 2021 considerably higher temporal resolutions up to nanoseconds for 2021 are available, but anticipating later findings, we point out that the non--Markovian effects are relevant on time scales much larger than one second.
We use the midpoint price
\begin{align*}
	m_i(t) = \frac{a_i(t)+b_i(t)}{2}
\end{align*}
as average of the best as $a_i(t)$ and best bid $b_i(t)$ for stock $i$ in the order book where $t$ runs in one--second steps. 
We calculate all the midpoint prices, but use only the last one for each second in the further calculations.
We introduce the arithmetic returns in the midpoint prices
\begin{equation}
	r_i(t,\tau) = \frac{m_i(t+\tau) - m_i(t)}{m_i(t)}
\end{equation}
with a time lag $\tau$. Here, we use the midpoint prices instead of the trade prices as we are also interested in the dynamics of the order book near the midpoint between consecutive trades.
As in Ref.~\cite{WSG2015preprint,Wang2016,Wang2016_2}, we define the trading sign $\varepsilon_i(t; n)$ of a single trade $n$ by the sign of the difference between the trade price $S_i(t; n)$ and the previous trade price $S_i(t; n-1)$
\begin{align}       
	\varepsilon_i(t;n)=\begin{cases}
		\mathrm{sgn}\bigl(S_i(t;n)-S_i(t;n-1)\bigr), & \text{if $S_i(t;n)\neq S_i(t;n-1)$}\\
		\varepsilon_i(t;n-1), & \text{otherwise}
	\end{cases}   \,,     
\end{align}
where $n$ counts the trades within the given time window $t$ of length one second.

If the price rises between two trades, the trading sign is $\varepsilon_i(t;n) = +1$.
An increase in the price can result from a sellout of the entire volume at the best ask $a_i(t;n-1)$.
Hence a trade sign $\varepsilon(t;n) = +1$ indicates that a trade was triggered by a buy order placed at time $(t; n-1)$. Similarly, a trade sign $\varepsilon(t;n) = -1$ means that a trade was triggered by a sell order at the time $(t; n-1)$.
To obtain a trade sign for whole interval $t$, we sum up all $N(t)$ trade signs within a second
\begin{align}       
	\varepsilon_i(t)=\left\{                  
	\begin{array}{lll}    
		\mathrm{Sgn}
		\left(\sum\limits_{n=1}^{N(t)}\varepsilon_i(t;n)\right) & \ , \quad & \mbox{if} \quad N(t)>0  \\    
		0 & \ , \quad & \mbox{if} \quad N(t)= 0 
	\end{array}    \,.       
	\right.
	\label{eqn:tradeSignSecond}              
\end{align} 
with
\begin{align}       
	\mathrm{Sgn}(x)=\left\{                  
	\begin{array}{lll}    
		+1 & \ , \quad & 	\quad x >0  \\    
		\hspace{0.27cm}0 & \ , \quad &  \quad x = 0  \\
		-1 & \ , \quad & 	\quad x < 0  \\    
	\end{array}    \,.       
	\right.
	\label{eqn:Sgn}              
\end{align}
A value of zero for $\varepsilon_i(t)$ either indicates that the same quantity of market orders to buy or sell cancel each other out for $N(t)>0$, \textit{i.\,e.} there is a balance of buy and sell market orders,
or no trading has taken place during a second for $N(t)=0$. We emphasize that no trading or delays in trading can also be part of a trading strategy. Excluding the trade sign zero makes the analysis more similar to studying markets on a trading time scale.
The accuracy of Eq.~\eqref{eqn:tradeSignSecond} was validated in Ref.~\cite{Wang2016}.

We measure the response of a trade in stock $j$ at time $t$ on the stock $i$ at a later time $t+\tau$
by the response functions
\begin{equation}
	R_{ij}(\tau) = \langle r_i(t,\tau)\varepsilon_j(t) \rangle_t - \langle r_i(t,\tau) \rangle_t \langle \varepsilon_j(t)\rangle _t \,.
	\label{eqn:crossResponseFunction}   
\end{equation}
For $i=j$, we obtain the self--response function and for $i \neq j$ the cross--response function.
From a strict mathematics viewpoint, we deal with time--lagged covariances of two time series. However, as the two time series are of different kind, an indicator function $\varepsilon_j(t)$ and an ordinary time series $r_i(t,\tau)$, we refer to \eqref{eqn:crossResponseFunction} as response functions. Since the average of the trade sign self--correlator $\langle \varepsilon_j(t)\rangle _t$ is close to zero for financial markets, definitions of response function without the second term are also common.
The trade sign correlator
\begin{equation}
	\Theta_{ij}(\tau) = \langle \varepsilon_i(t+\tau)\varepsilon_j(t) \rangle _t - \langle \varepsilon_i(t+\tau) \rangle _t \langle \varepsilon_j(t) \rangle _t 
	\label{eqn:TradeSignCorrelator}   
\end{equation}
quantifies time--lagged covariances in the trade sign time series.
We obtain the trade sign self--correlator for $i = j$ and the trade sign cross--correlator for $i \neq j$.  %
Only when zero trade signs are not included, Eq.~\eqref{eqn:TradeSignCorrelator} is a time--lagged correlation as the time series of trade signs then have unit variance.
Analogous to the response function, the cross--correlator $\Theta_{ij}(\tau)$ is often written without the second term.
Finally, we average the cross--response functions and cross--correlators over all trading days for a stock or a stock pair $i,j$ and a given time lag $\tau$.

\section{\label{sec:Results}Results}

In Secs.~\ref{sec:MarketResponseFunctions} and~\ref{sec:TradeSignCorrelators}, we work out the responses and trade sign correlations for the whole market by averaging across all individual stocks and stock pairs, respectively. 
In Sec.~\ref{sec:InterIntraSectorResponse}, we decompose the cross--response functions into intra-- and intersectoral responses.
We visualize the responses for the market as a whole in Sec.~\ref{sec:VisualisationResponses}. Finally, we introduce and compare active and passive responses for the different years in Sec.~\ref{sec:ActivePassiveResponseFunctions}.

\subsection{\label{sec:MarketResponseFunctions}Market Response Functions}

\begin{figure}[htbp]
	\captionsetup[subfigure]{labelformat=empty}
	\centering
	\begin{minipage}{0.5\textwidth}
		\subfloat[]{\begin{overpic}[width=1.0\linewidth]{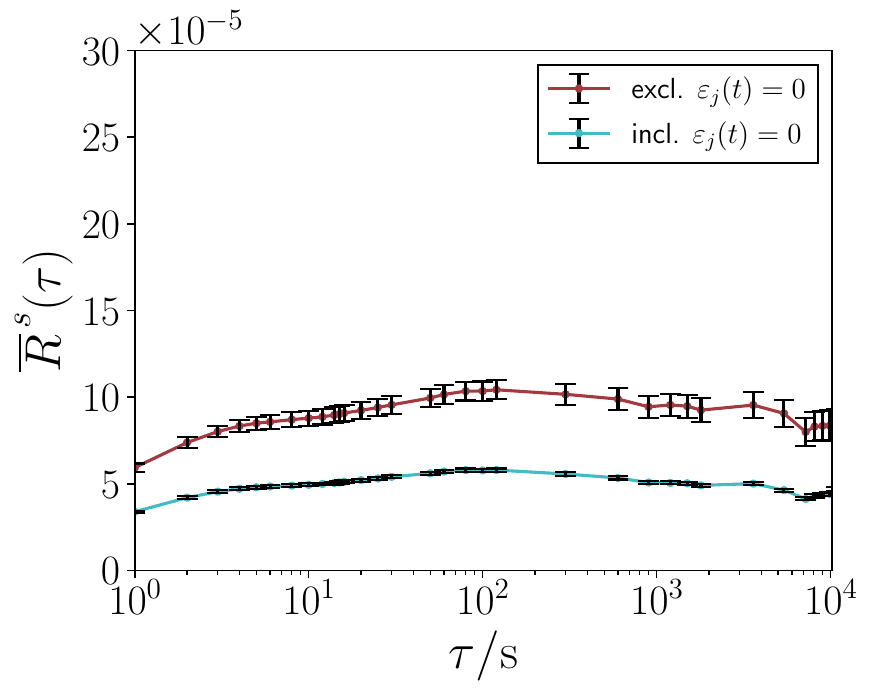}
				\put(19,61){\noindent\fbox{\parbox{0.8cm}{\sffamily 2007}}}
		\end{overpic}}%
	\end{minipage}%
	\begin{minipage}{0.5\textwidth}
		\subfloat[]{\begin{overpic}[width=1.0\linewidth]{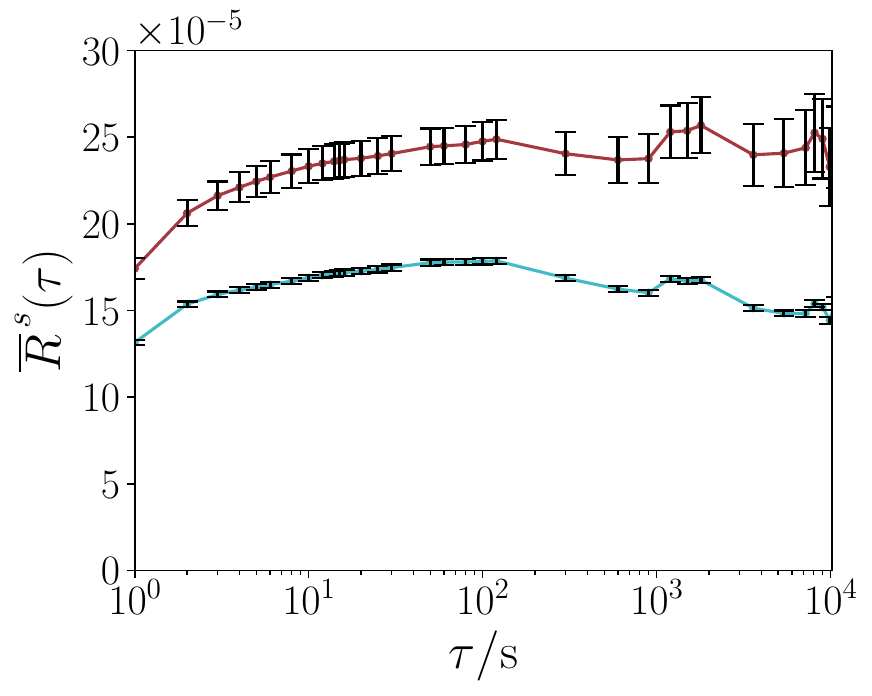}
				\put(19,31){\noindent\fbox{\parbox{0.8cm}{\sffamily 2008}}}
		\end{overpic}}%
	\end{minipage}
	\begin{minipage}{0.5\textwidth}
		\subfloat[]{\begin{overpic}[width=1.0\linewidth]{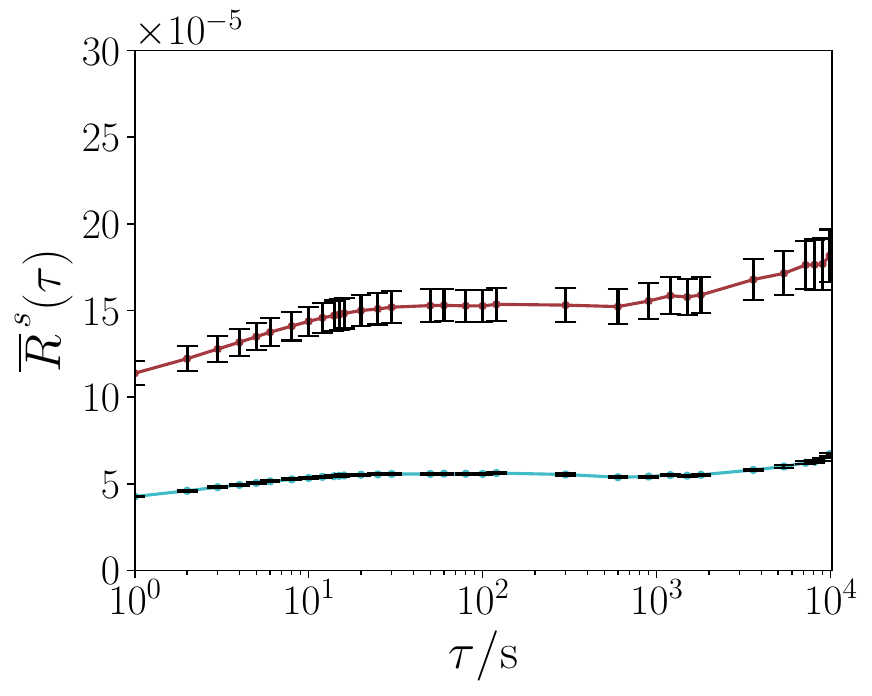}
				\put(19,61){\noindent\fbox{\parbox{0.8cm}{\sffamily 2014}}}
		\end{overpic}}%
	\end{minipage}%
	\begin{minipage}{0.5\textwidth}
		\subfloat[]{\begin{overpic}[width=1.0\linewidth]{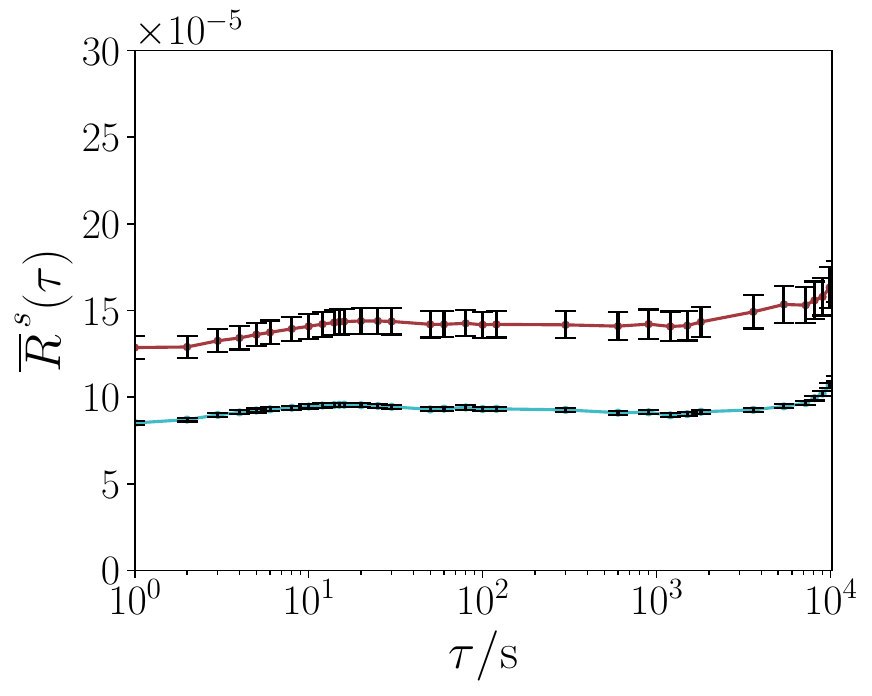}
				\put(19,61){\noindent\fbox{\parbox{0.8cm}{\sffamily 2021}}}
		\end{overpic}}%
	\end{minipage}
	\caption{\label{fig:MarketResponse_Self}Market self--responses for 2007, 2008, 2014 and 2021 versus time lag $\tau$. Trade signs $\varepsilon_i(t) = 0$ are excluded/included in the red/blue curves.
		For better comparison,
		the response functions including $\varepsilon_j(t)=0$ are multiplied by a factor of 6. The error bars indicate deviations for all stocks for a given time lag $\tau$ and are not multiplied by a factor.}
\end{figure}

We obtain a better statistical significance for the estimation of market self-- and cross--impact by averaging across all self--responses
\begin{equation}
	\overline{R}^{\,s}(\tau) \ = \ \langle\langle R_{ii}(\tau)\rangle _{i}\rangle _{i}
	\label{eqn:SelfAveragedResponse}
\end{equation}
and across all cross-responses
\begin{equation}
	\overline{R}^{\,c}(\tau) \ = \ \langle\langle R_{ij}(\tau)\rangle _{j}\rangle _{i} \;,  \quad \text{for $i \neq j$}
	\label{eqn:doubleAveragedResponse}
\end{equation}
for a given time lag $\tau$. Averaging either over the elements for $i=j$ or for $i \neq j$ in Eq.~\eqref{eqn:doubleAveragedResponse} yields the self--response or the cross--response for the market to which we refer as market self--response or market cross--response, respectively.
Figure~\ref{fig:MarketResponse_Self} depicts market self--responses for 2007, 2008, 2014 and 2021. 
The market self--responses are always positive. Response values including trade sign zero are smaller than responses excluding trade sign zero, but the trend of the market cross--response functions is not affected.
For small time lags $\tau$, all curves  increase up to a time lag of around 100 seconds.
The error bars indicate the standard deviations after averaging over all 99 individual response functions for a given time lag $\tau$. We note that the error bars do not contain the information on the deviations in the individual responses for a whole trading year. Rather, the error bars show that the average of the individual self--responses do not differ much from each other.
Most strikingly, the market self--responses are not decreasing for large $\tau$ in 2014 and 2021, they even increase for very large time lags $\tau$.
The larger responses in 2008 might be explained by %
a higher risk--aversion of the traders due to the financial crisis of 2008/2009. Put differently, liquidity providers demanded a higher risk compensation or risk premium for a potential trade.
For 2014 and 2021, the market self--responses excluding trade sign zero stay on a high level, \textit{i.e.} the results indicate that the crisis had a lasting influence on the traders' risk aversion.

\begin{figure}[htbp]
	\captionsetup[subfigure]{labelformat=empty}
	\centering
\begin{minipage}{0.5\textwidth}
	\subfloat[]{\begin{overpic}[width=0.9\linewidth]{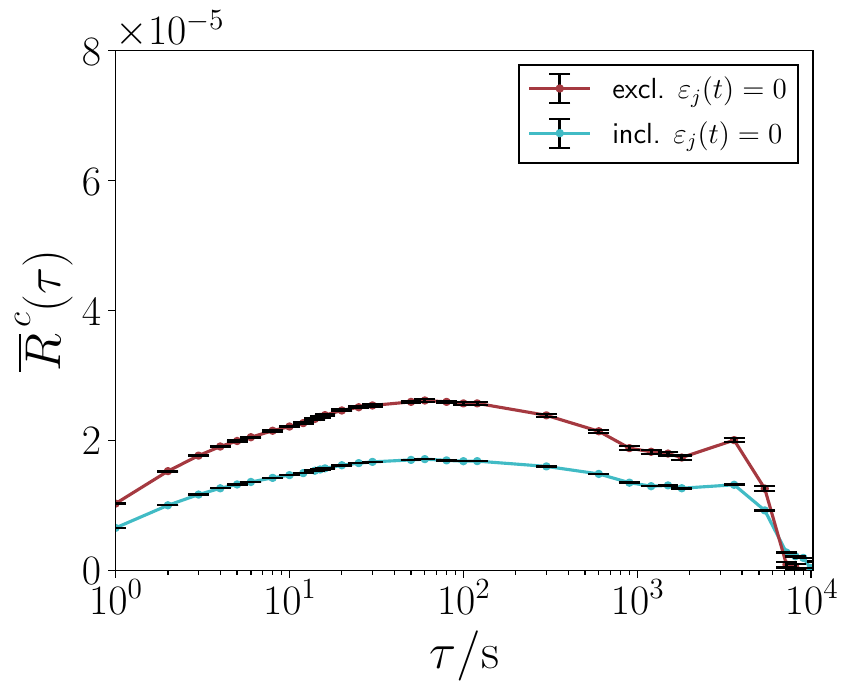}
			\put(19,61){\noindent\fbox{\parbox{0.8cm}{\sffamily 2007}}}
	\end{overpic}}%
\end{minipage}%
\begin{minipage}{0.5\textwidth}
	\subfloat[]{\begin{overpic}[width=0.9\linewidth]{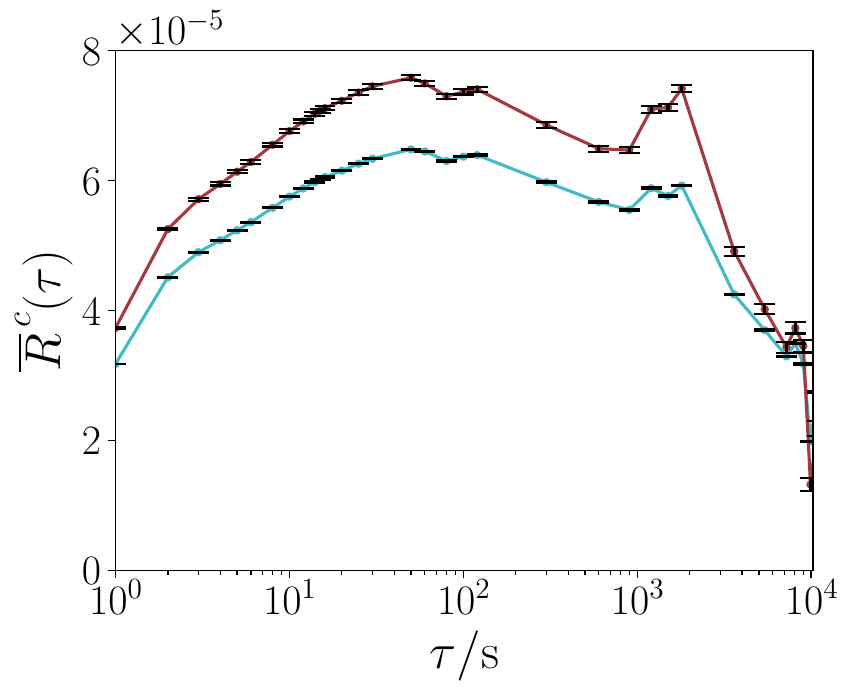}
			\put(19,31){\noindent\fbox{\parbox{0.8cm}{\sffamily 2008}}}
	\end{overpic}}%
\end{minipage}
	\begin{minipage}{0.5\textwidth}
		\subfloat[]{\begin{overpic}[width=0.9\linewidth]{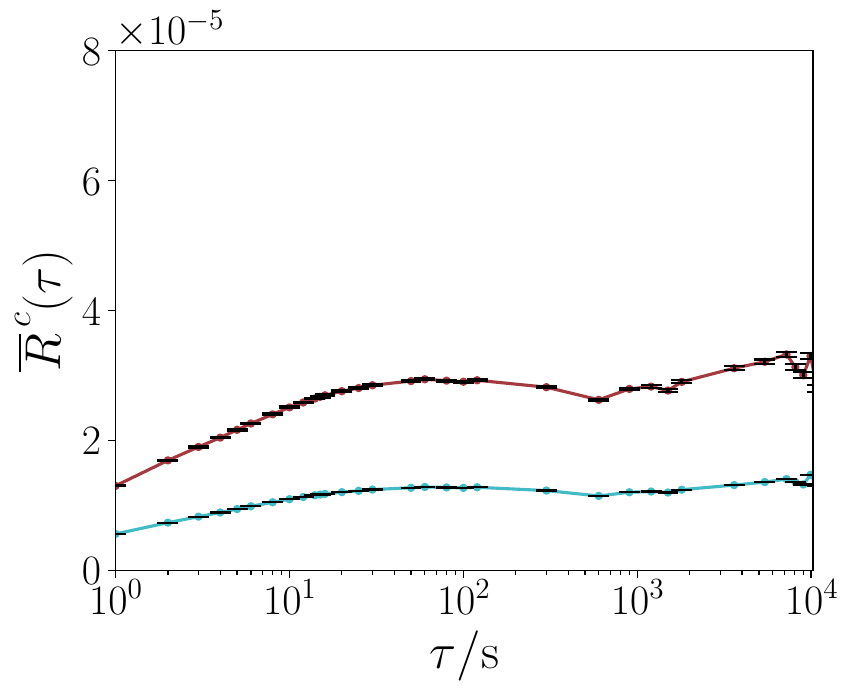}
				\put(19,61){\noindent\fbox{\parbox{0.8cm}{\sffamily 2014}}}
		\end{overpic}}%
	\end{minipage}%
	\begin{minipage}{0.5\textwidth}
		\subfloat[]{\begin{overpic}[width=0.9\linewidth]{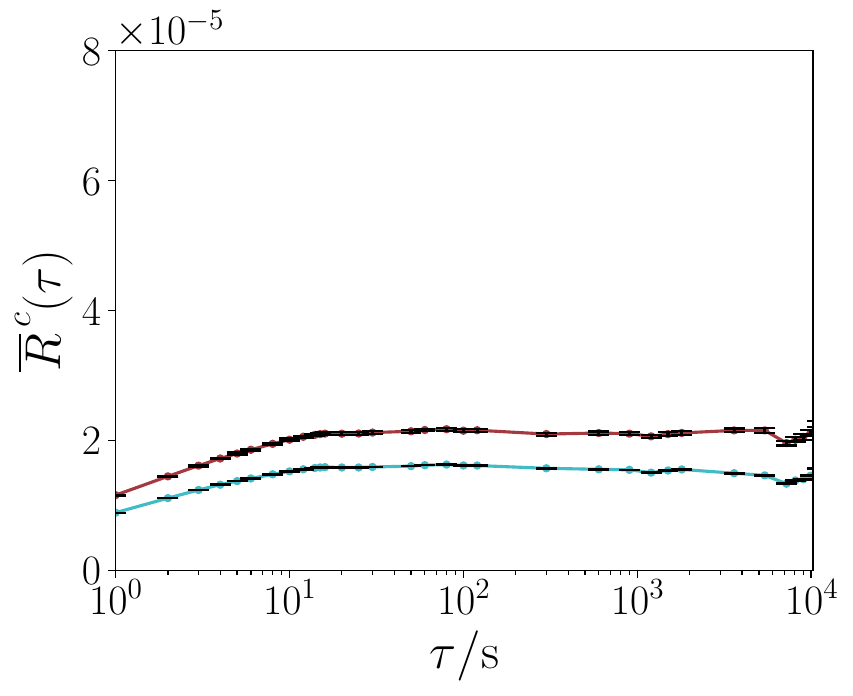}
				\put(19,61){\noindent\fbox{\parbox{0.8cm}{\sffamily 2021}}}
		\end{overpic}}%
	\end{minipage}
	\caption{\label{fig:MarketResponse}Market cross--responses for 2007, 2008, 2014 and 2021 versus time lag $\tau$. Trade signs $\varepsilon_i(t) = 0$ are excluded/included in the red/blue curves.
	For better comparison,
		the response functions including $\varepsilon_j(t)=0$ are multiplied by a factor of 6. The error bars indicate deviations for all stock pairs for a given time lag $\tau$ and are not multiplied by a factor.
	 Results for 2008 are replicated from Ref.~\cite{Wang2016}.}
\end{figure}

Figure~\ref{fig:MarketResponse} shows the market cross-responses for 2007, 2008, 2014 and 2021.
The market self--responses are always larger than the market cross--response function in Fig.~\ref{fig:MarketResponse_Sectors}.
Differences between the years are clearly discernable.
The market cross--responses decrease for 2007 and 2008 for larger time lags $\tau$, while this is not the case in 2014 and 2021 for larger time lags $\tau$. 
The largest values for the market response are measured in the crisis year 2008. 
In contrast to the market self--responses for 2014 and 2021, the cross--responses return to similar values as in 2007.
As observed in Ref.~\cite{Wang2016}, responses including trade sign zero are smaller than responses excluding trade sign zero, but the trend of the market cross-response functions is not affected.

\subsection{\label{sec:InterIntraSectorResponse}Inter- and Intrasectoral Cross--Response}

\begin{figure}[htbp]
	\captionsetup[subfigure]{labelformat=empty}
	\centering
	\begin{minipage}{0.5\textwidth}
		\subfloat[]{\begin{overpic}[width=0.9\linewidth]{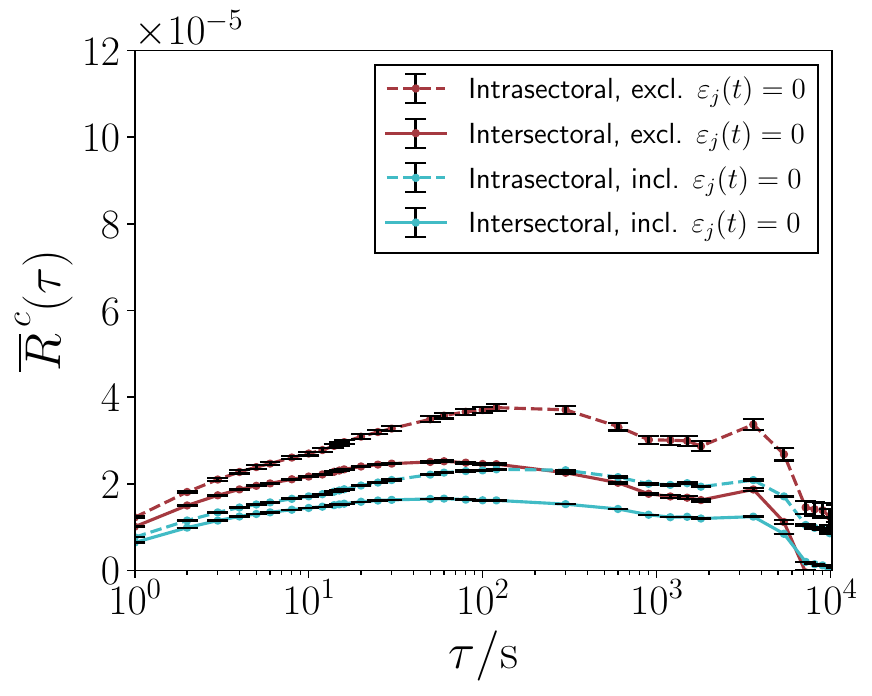}
				\put(19,61){\noindent\fbox{\parbox{0.8cm}{\sffamily 2007}}}
		\end{overpic}}%
	\end{minipage}%
	\begin{minipage}{0.5\textwidth}
		\subfloat[]{\begin{overpic}[width=0.9\linewidth]{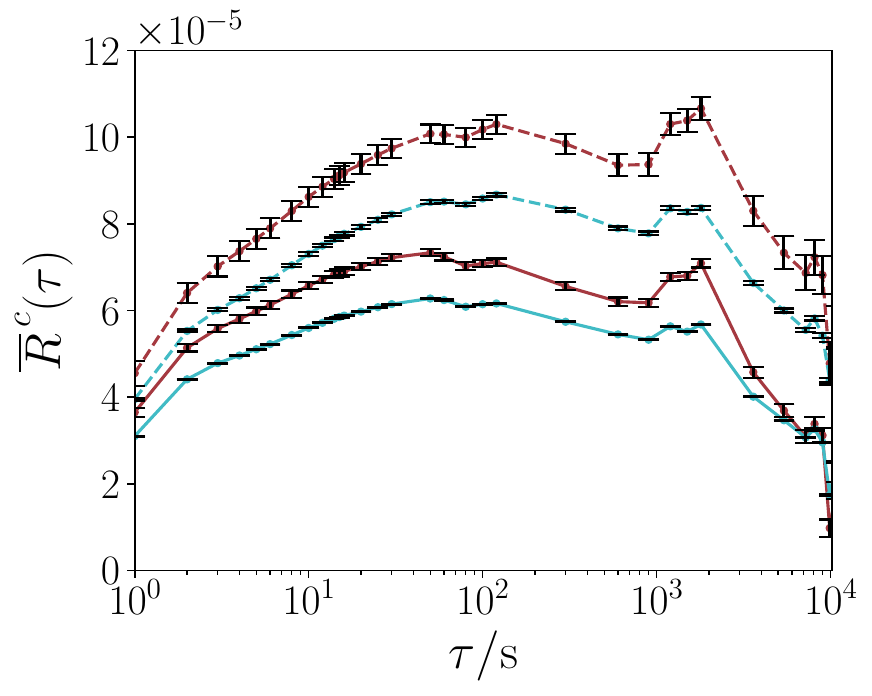}
				\put(19,27){\noindent\fbox{\parbox{0.8cm}{\sffamily 2008}}}
		\end{overpic}}%
	\end{minipage}
	\begin{minipage}{0.5\textwidth}
		\subfloat[]{\begin{overpic}[width=0.9\linewidth]{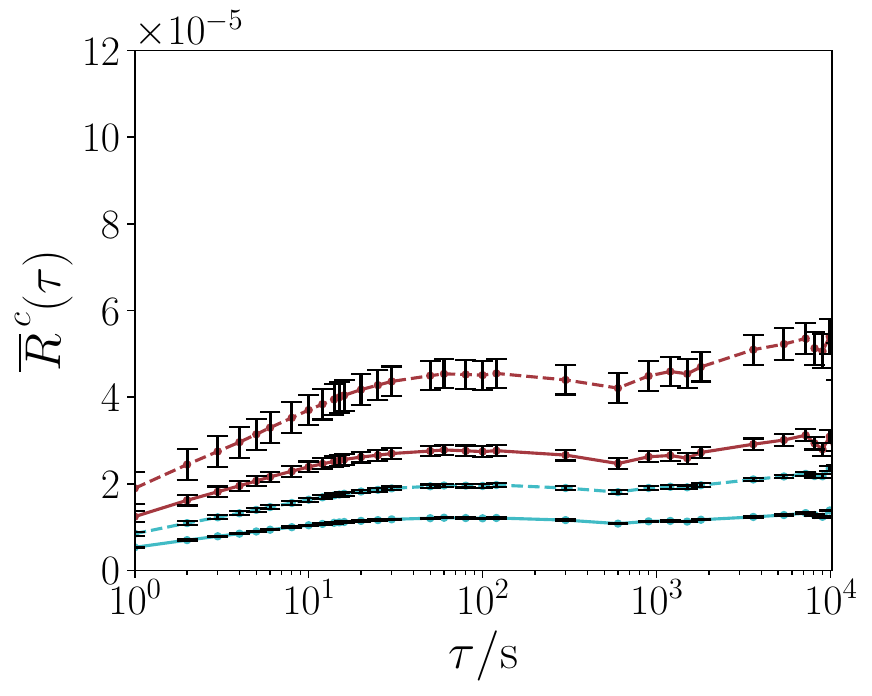}
				\put(19,61){\noindent\fbox{\parbox{0.8cm}{\sffamily 2014}}}
		\end{overpic}}%
	\end{minipage}%
	\begin{minipage}{0.5\textwidth}
		\subfloat[]{\begin{overpic}[width=0.9\linewidth]{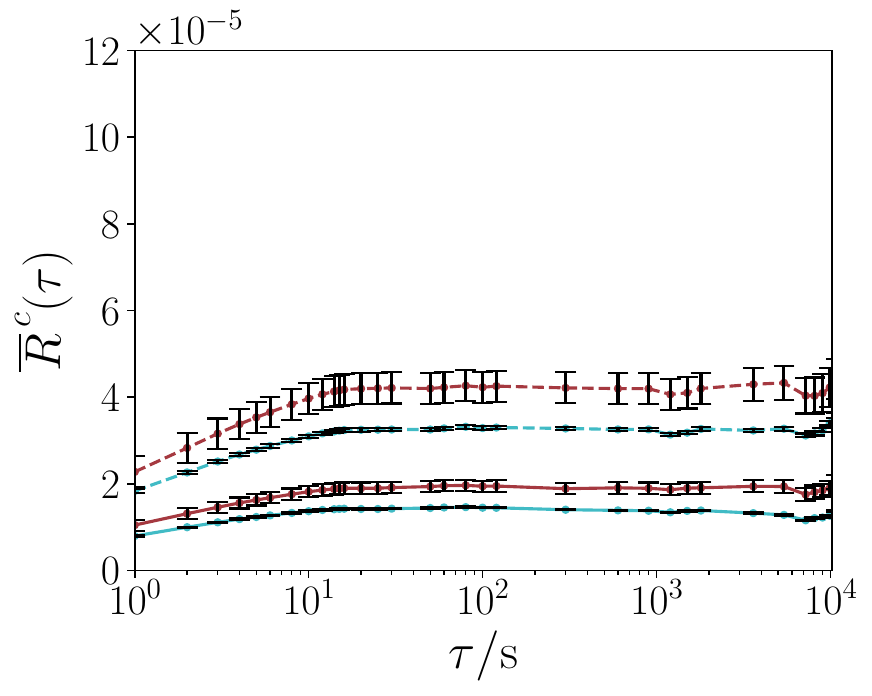}
				\put(19,61){\noindent\fbox{\parbox{0.8cm}{\sffamily 2021}}}
		\end{overpic}}%
	\end{minipage}
	\caption{\label{fig:MarketResponse_Sectors}Market cross--responses for 2007, 2008, 2014 and 2021 versus time lag $\tau$. Trade signs $\varepsilon_i(t) = 0$ are excluded/included in the red/blue curves.
		For better comparison,
		the response functions including $\varepsilon_j(t)=0$ are multiplied by a factor of 6. The error bars indicate deviations for all stock pairs for a given time lag $\tau$ and are not multiplied by a factor.}
\end{figure}

We study common trends of cross--responses for all stock pairs within and between all ten industrial sectors, see Sec.~\ref{sec:DataSet}. We work out their averages to which we refer as intersectoral and intrasectoral cross--response, respectively, shown in Fig.~\ref{fig:MarketResponse_Sectors}.
The intrasectoral cross--responses are usually larger than the intersectoral ones. 
This presumably reflects the strong intrasectoral and the weaker intersectoral correlations which are visible in the striking block structure of the Pearson correlation matrix.
The observations in Sec.~\ref{sec:MarketResponseFunctions} also apply to inter-- and intrasectoral cross--responses.

\subsection{\label{sec:TradeSignCorrelators}{Trade Sign Correlators}}

\begin{figure}[htbp]
	\captionsetup[subfigure]{labelformat=empty}
	\centering
	\begin{minipage}{0.5\textwidth}
		\subfloat[]{\begin{overpic}[width=1.0\linewidth]{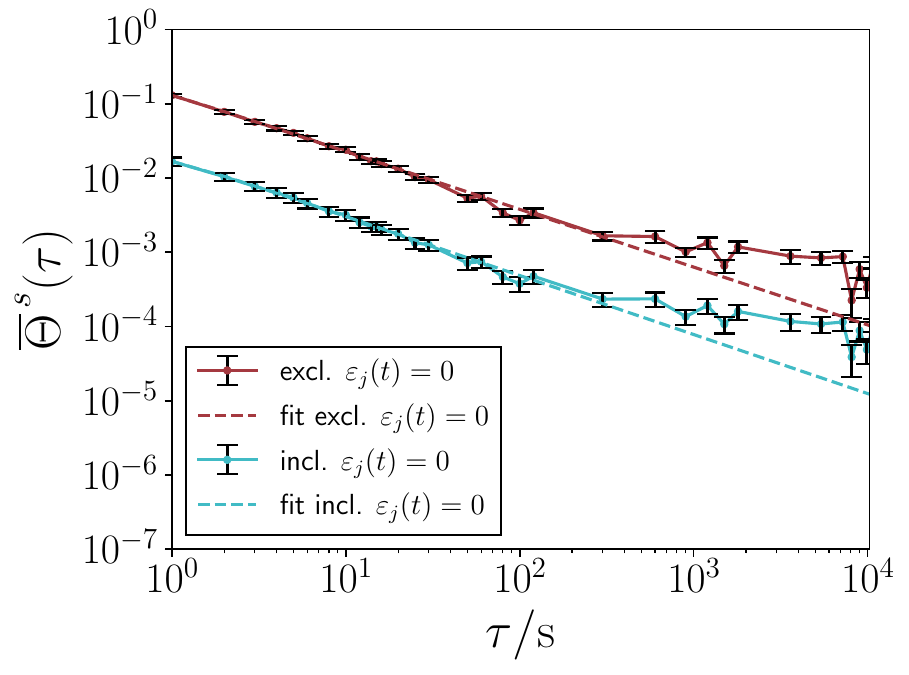}
				\put(76,61){\noindent\fbox{\parbox{0.8cm}{\sffamily 2007}}}
		\end{overpic}}%
	\end{minipage}%
	\begin{minipage}{0.5\textwidth}
		\subfloat[]{\begin{overpic}[width=1.0\linewidth]{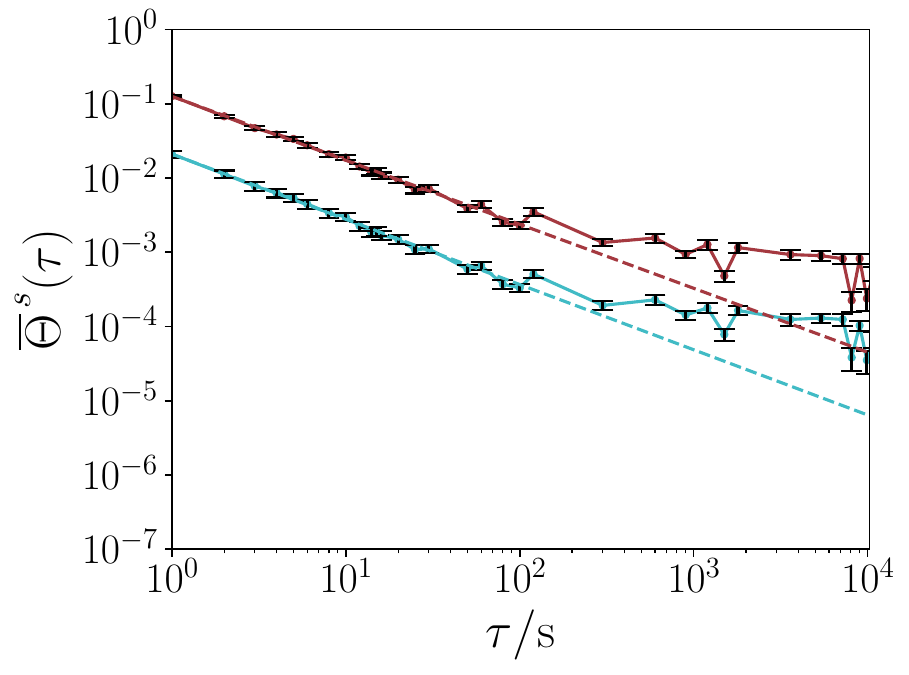}
				\put(76,61){\noindent\fbox{\parbox{0.8cm}{\sffamily 2008}}}
		\end{overpic}}%
	\end{minipage}
	\begin{minipage}{0.5\textwidth}
		\subfloat[]{\begin{overpic}[width=1.0\linewidth]{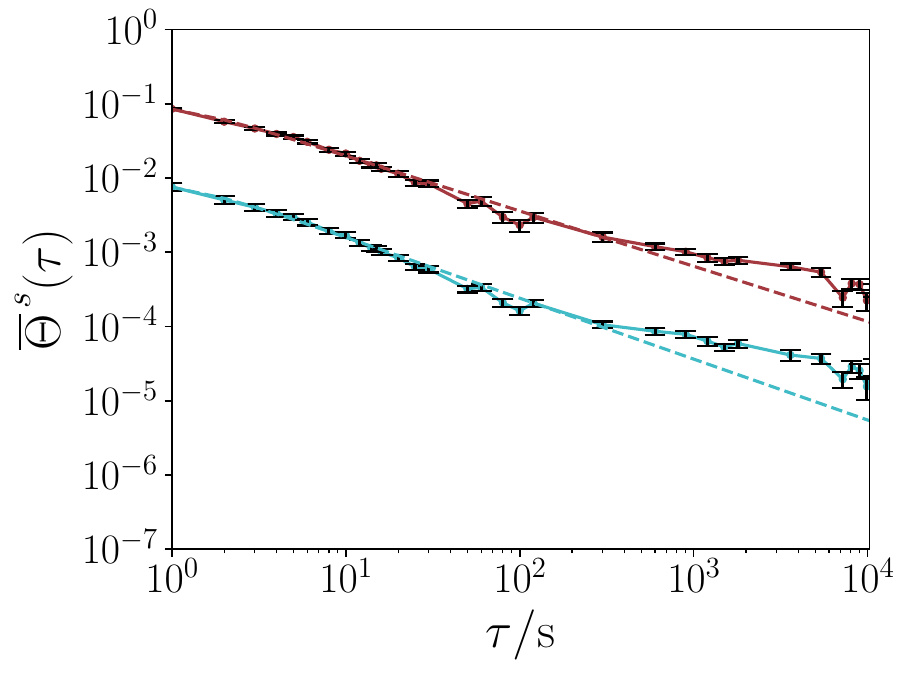}
				\put(76,61){\noindent\fbox{\parbox{0.8cm}{\sffamily 2014}}}
		\end{overpic}}%
	\end{minipage}%
	\begin{minipage}{0.5\textwidth}
		\subfloat[]{\begin{overpic}[width=1.0\linewidth]{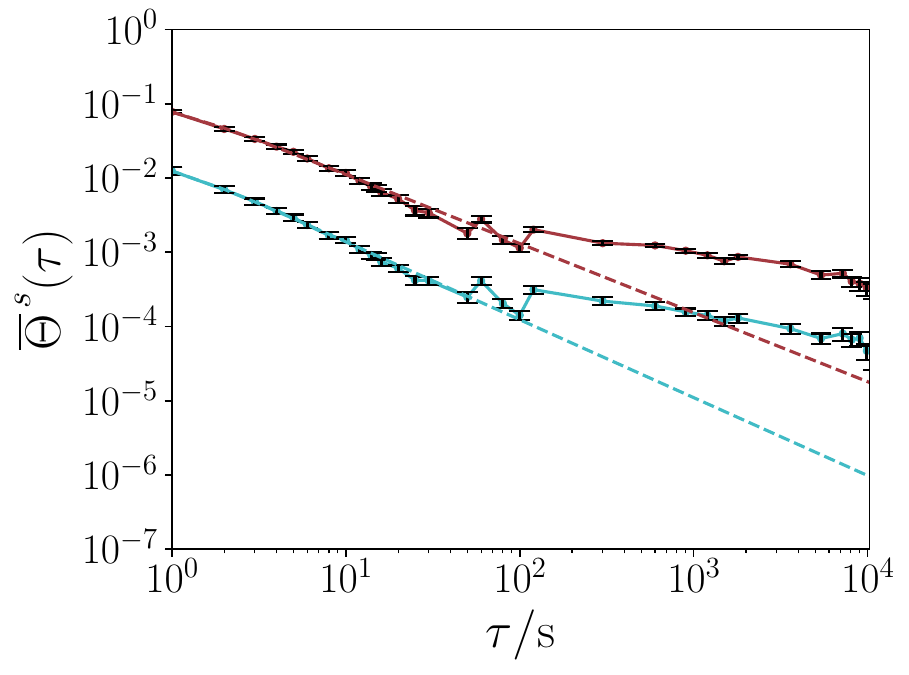}
				\put(76,61){\noindent\fbox{\parbox{0.8cm}{\sffamily 2021}}}
		\end{overpic}}%
	\end{minipage}
	\caption{\label{fig:Market_TradeSign_Self}Market trade sign self--correlator for 2007, 2008, 2014 and 2021 versus time lag $\tau$, shown together with power law fits. Trade signs $\varepsilon_i(t) = 0$ are excluded/included in the red/blue curves. The error bars indicate deviations for all stocks for a given time lag $\tau$.}
\end{figure}

\begin{figure}[htbp]
	\captionsetup[subfigure]{labelformat=empty}
	\centering
	\begin{minipage}{0.5\textwidth}
		\subfloat[]{\begin{overpic}[width=1.0\linewidth]{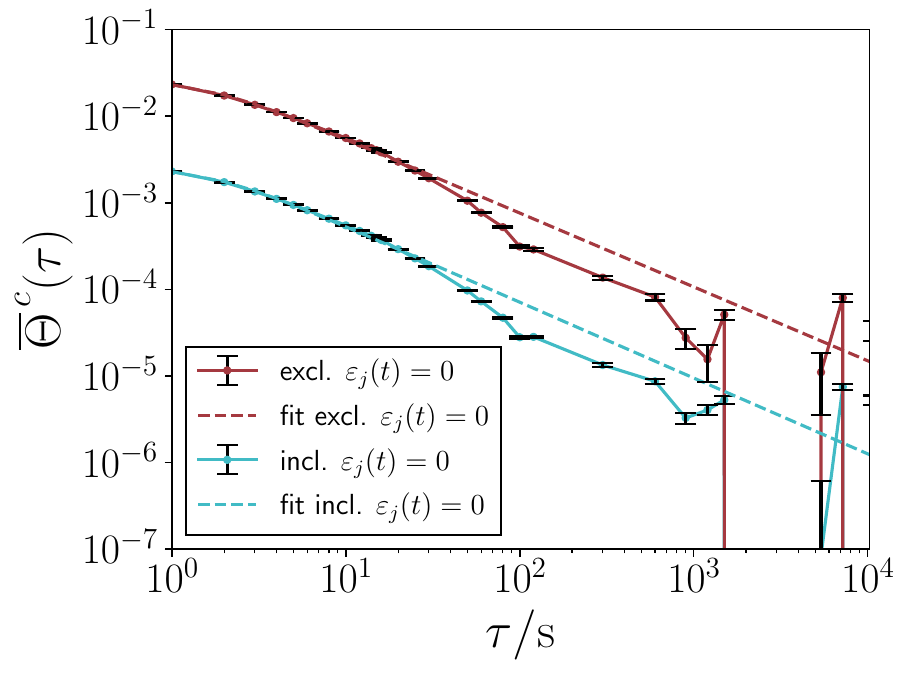}
				\put(76,61){\noindent\fbox{\parbox{0.8cm}{\sffamily 2007}}}
		\end{overpic}}%
	\end{minipage}%
	\begin{minipage}{0.5\textwidth}
		\subfloat[]{\begin{overpic}[width=1.0\linewidth]{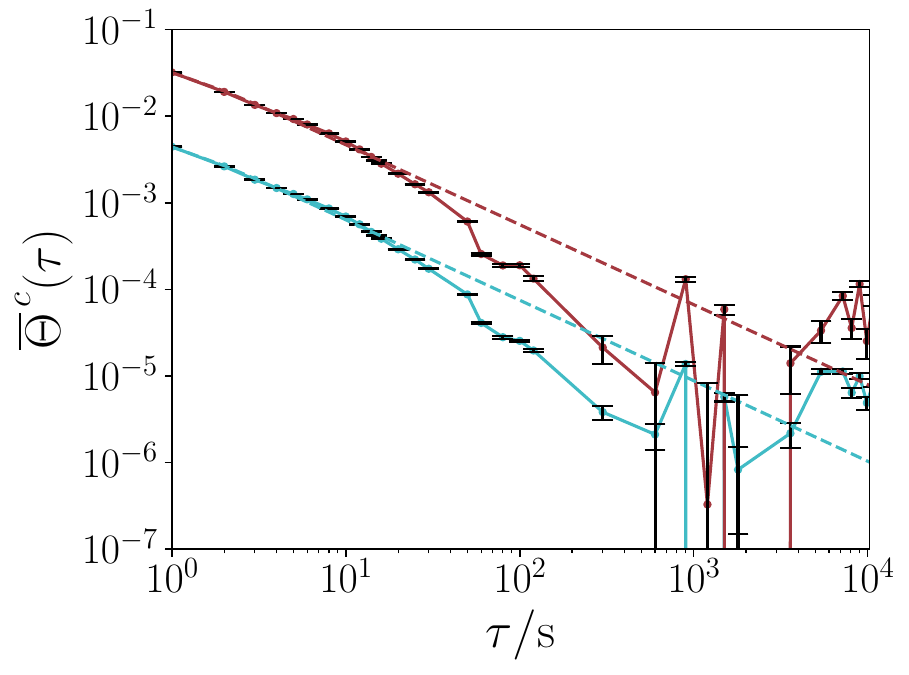}
				\put(76,61){\noindent\fbox{\parbox{0.8cm}{\sffamily 2008}}}
		\end{overpic}}%
	\end{minipage}
	\begin{minipage}{0.5\textwidth}
		\subfloat[]{\begin{overpic}[width=1.0\linewidth]{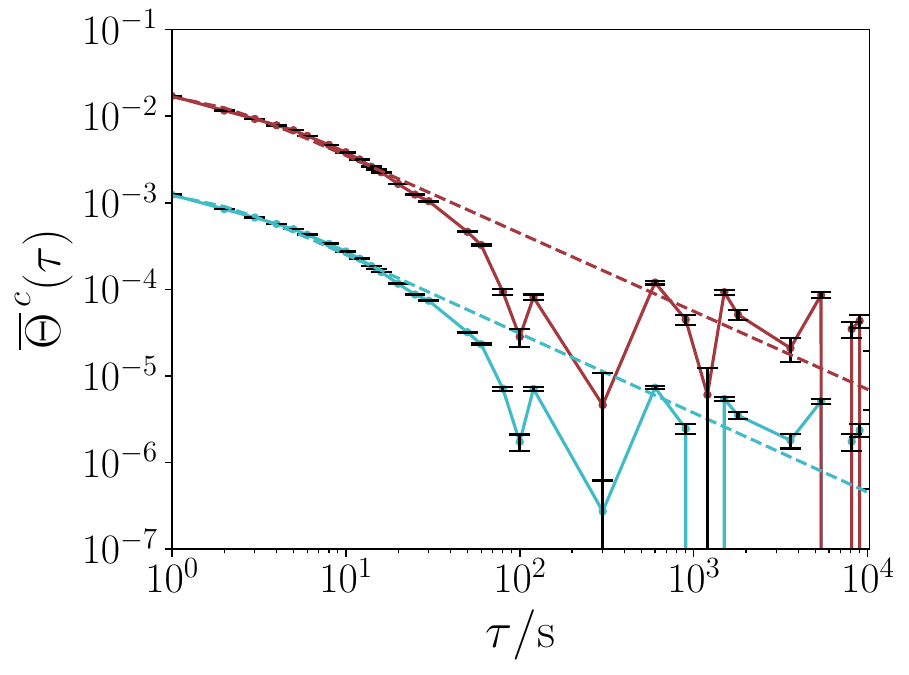}
				\put(76,61){\noindent\fbox{\parbox{0.8cm}{\sffamily 2014}}}
		\end{overpic}}%
	\end{minipage}%
	\begin{minipage}{0.5\textwidth}
		\subfloat[]{\begin{overpic}[width=1.0\linewidth]{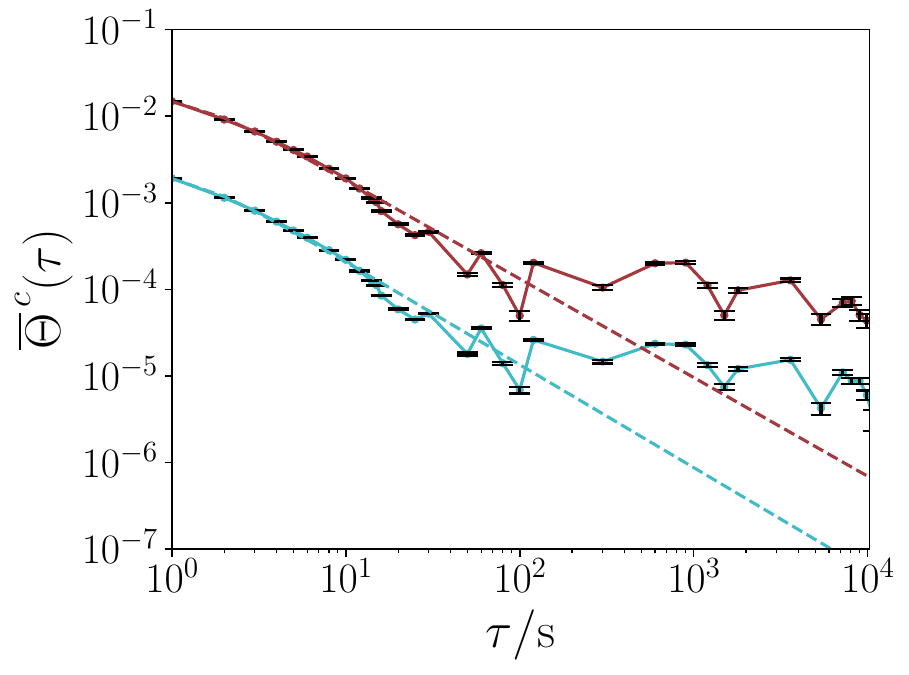}
				\put(76,61){\noindent\fbox{\parbox{0.8cm}{\sffamily 2021}}}
		\end{overpic}}%
	\end{minipage}
	\caption{\label{fig:Market_TradeSign}Market trade sign cross--correlator for 2007, 2008, 2014 and 2021 versus time lag $\tau$, shown together with power law fits. Trade signs $\varepsilon_i(t) = 0$ are excluded/included in the red/blue curves. The error bars indicate deviations for all stock pairs for a given time lag $\tau$.}
\end{figure}

Analogously to the market response function in Eq.~\eqref{eqn:doubleAveragedResponse} we study the trade sign self--correlator
\begin{equation}
	\overline{\Theta}^{\,s}(\tau) \ = \ \langle\langle \Theta_{ii}(\tau) \rangle _{i}\rangle _{i}  
\end{equation}
by averaging over all elements for $i = j$ and the trade sign cross--correlator
\begin{equation}
	\overline{\Theta}^{\,c}(\tau) \ = \ \langle\langle \Theta_{ij}(\tau) \rangle _{j}\rangle _{i} \;, \quad \text{for $i \neq j$}  \ , 
\end{equation}
as displayed in Figs.~\ref{fig:Market_TradeSign_Self} and \ref{fig:Market_TradeSign}.
As the inclusion of the trade sign zero contributes to negative values of the trade sign, the values decrease accordingly, see~Ref.~\cite{Wang2016}.
The power law trend
is fitted very well by the function
\begin{equation}
	\overline{\Theta}^{\,{\kappa}}(\tau)  =\frac{\overline{\vartheta}^{\,{\kappa}} }{\left(1+(\tau/\overline{\tau}^{\,{\kappa}} )^2\right)^{\overline{\gamma}^{\,\kappa}/2}} \ , \quad \kappa = s, c \,.
	\label{eq34}
\end{equation}
We list the fit parameters $\overline{\vartheta}^{\,{\kappa}}$,  
$\overline{\tau}^{\,{\kappa}}$ and $\overline{\gamma}^{\,{\kappa}}$, $\kappa = s ,c$, in Tab.~\ref{tab:averagedFitParameters}.
The two--fold averaging improves the statistics especially for larger values of time lag $\tau$.
The results for $\overline{\gamma}^{\,{s}}$ and $\overline{\gamma}^{\,{c}}$ are interesting as values above one indicate short--term memory and values smaller than one long--term memory~\cite{Beran2017}. 
For 2007, we notice values smaller than 0.8, for the other years they are larger except for the year 2014 when excluding trade sign zero.
For 2021, the values $\overline{\gamma}^{\,s}$ and $\overline{\gamma}^{\,c}$ even take larger values than one apart from the case when excluding zero.

Furthermore, we notice for the trade sign self--correlator a change in the power law between 100\,s and 1000\,s, \textit{i.e.} we find that a second, distinct regime emerges for larger time lags $\tau$. This change in the power law is particularly evident for the year 2021 and indicates that the different traders' behavior changes
for larger time lags $\tau$ with a smaller parameter $\overline{\gamma}^{\,{s}}(\tau)$. For 2021, we see similar changes in the power law for the cross-correlator.

\begin{table}[htbp]
	\centering
	\caption{Fit parameters and normalized $\chi^2$~\cite{Wang2016} for the average trade sign cross--correlators.}
	\resizebox{\textwidth}{!}{%
		\begin{tabular}{@{}cccccccccc@{}}
			\toprule
			market sign correlator                        & year & \multicolumn{2}{c}{$\overline{\vartheta}^{\,{s}}, \overline{\vartheta}^{\,{c}}$} & \multicolumn{2}{c}{$\overline{\tau}^{\,{s}} [s], \overline{\tau}^{\,{c}} [s]$} & \multicolumn{2}{c}{$\overline{\gamma}^{\,{s}}, \overline{\gamma}^{\,{c}}$} & \multicolumn{2}{c}{$\chi^2 (\times 10^{-7})$} \\
			&      & incl. 0         & excl. 0         & incl. 0             & excl. 0             & incl. 0         & excl. 0         & incl. 0                  & excl. 0                 \\ \midrule
			\multirow{4}{*}{$\overline \Theta^{\,s}(\tau)$} & 2007 & 0.029           & 0.298           & 0.594               & 0.367               & 0.795           & 0.778           & 0.088                    & 4.544                   \\
			& 2008 & 4.496           & 30.425          & 0.002               & -0.001              & 0.878           & 0.862           & 0.154                    & 7.447                   \\
			& 2014 & 0.010           & 0.108           & 1.130               & 1.031               & 0.821           & 0.744           & 0.064                    & 10.067                  \\
			& 2021 & 0.022           & 0.128           & 0.725               & 0.742               & 1.050           & 0.930           & 0.088                    & 4.330                   \\ \midrule
			\multirow{4}{*}{$\overline \Theta^{\,c}(\tau)$} & 2007 & 0.003           & 0.027           & 1.611               & 1.530               & 0.875           & 0.853           & 0.004                    & 0.406                   \\
			& 2008 & 0.007           & 0.051           & 0.743               & 0.754               & 0.929           & 0.923           & 0.020                    & 1.032                   \\
			& 2014 & 0.001           & 0.020           & 1.517               & 1.478               & 0.916           & 0.901           & 0.004                    & 0.805                   \\
			& 2021 & 0.003           & 0.020           & 1.183               & 1.240               & 1.188           & 1.139           & 0.003                    & 0.326                   \\ \bottomrule
		\end{tabular}%
	\label{tab:averagedFitParameters}
	}
\end{table}

\subsection{\label{sec:VisualisationResponses}Visualization of Responses for the Market as a Whole}

We want to inspect the matrix structure of all self-- and cross--responses more closely. For a better visualization,
we define the normalized response functions,
\begin{equation}
	\rho_{ij}(\tau) \ = \ \frac{R_{ij}(\tau)}{\textrm{max\,}(|R_{ij}(\tau)|)} \,,
\end{equation}
which defines an asymmetric matrix $\rho$. On its diagonal are the normalized self--responses.
\begin{figure}[htbp]
	\centering
	\begin{minipage}{0.45\textwidth}
		{\includegraphics[width=1.\linewidth]{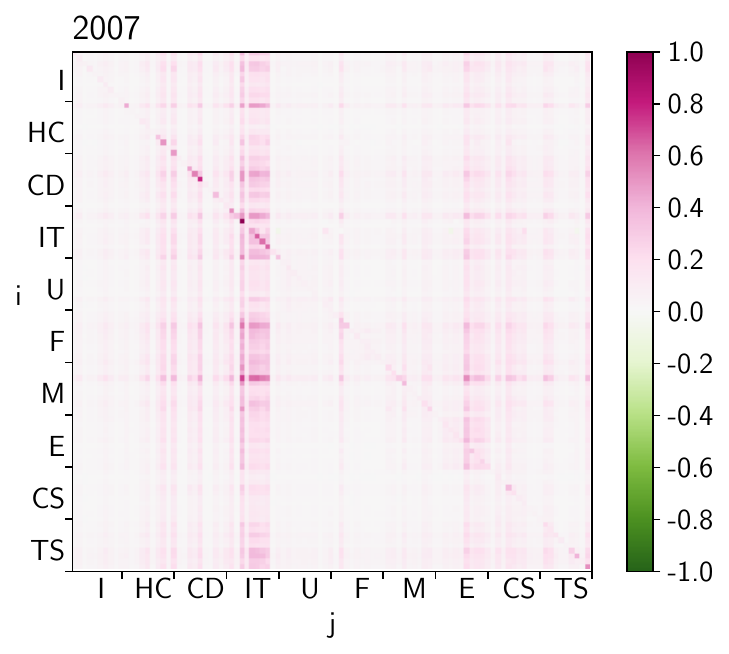}}%
	\end{minipage}%
	\begin{minipage}{0.45\textwidth}
		{\includegraphics[width=1.\linewidth]{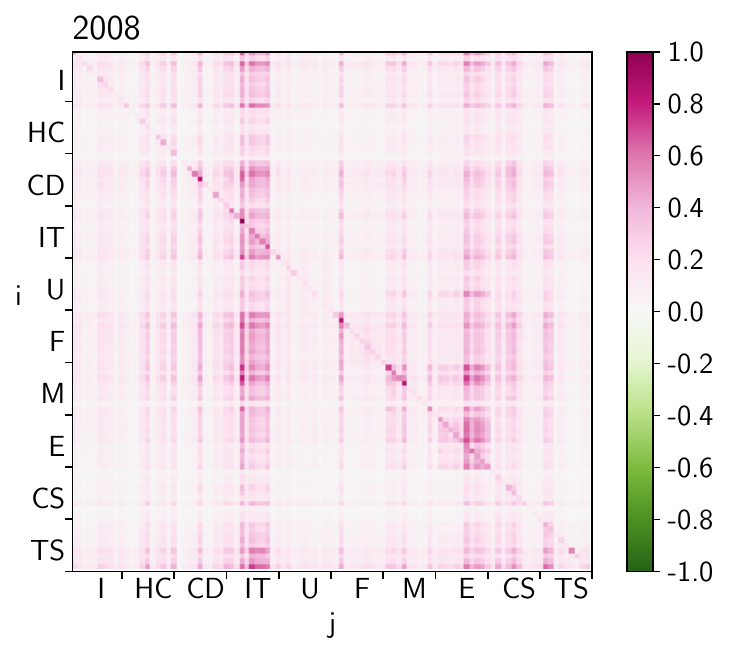}}%
	\end{minipage}
	\begin{minipage}{0.45\textwidth}
		{\includegraphics[width=1.\linewidth]{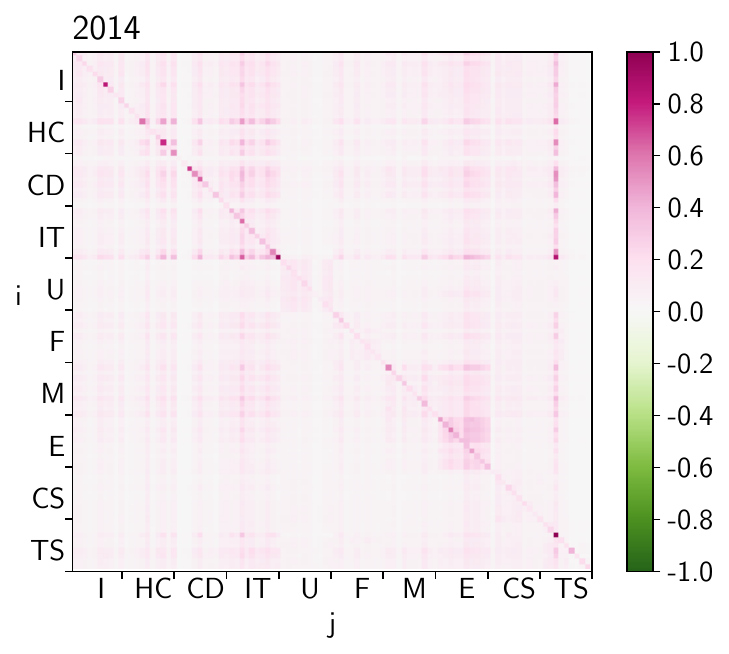}}%
	\end{minipage}%
	\begin{minipage}{0.45\textwidth}
		{\includegraphics[width=1.\linewidth]{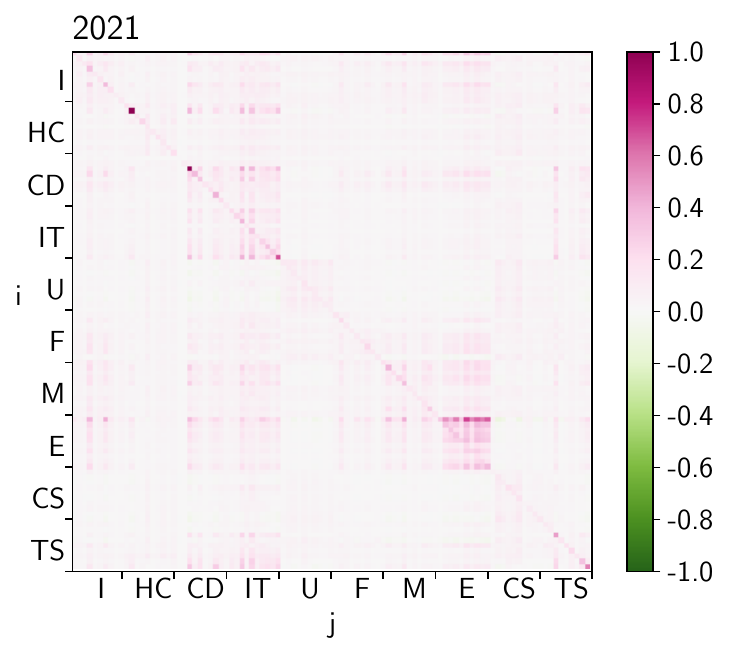}}%
	\end{minipage}%
	\caption{\label{fig:ResponsesMat}Matrices for the normalized responses \textit{including} $\varepsilon_i(t) = 0$ with entries $\rho_{ij}(\tau)$ for $i,j=1,\ldots,99$ at the time lag $\tau=30$~s in 2007, 2008, 2014 and 2021. The stocks pairs $(i,j)$ correspond to the sectors industrials (I), health care (HC), consumer discretionary (CD), information technology (IT), utilities (U), financials (F), materials (M), energy (E), consumer staples (CS), and telecommunications services (TS).}
\end{figure}
\begin{figure}[htbp]
	\centering
	\begin{minipage}{0.45\textwidth}
		{\includegraphics[width=1.\linewidth]{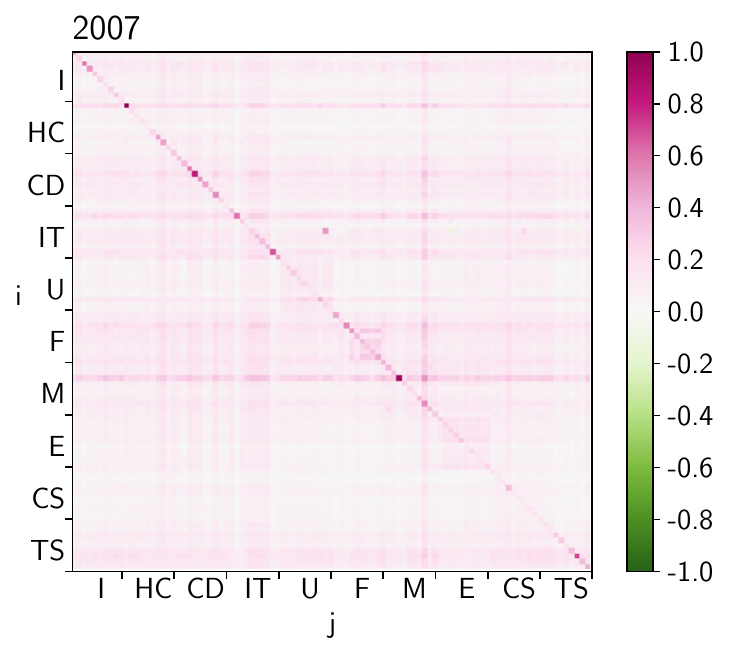}}%
	\end{minipage}%
	\begin{minipage}{0.45\textwidth}
		{\includegraphics[width=1.\linewidth]{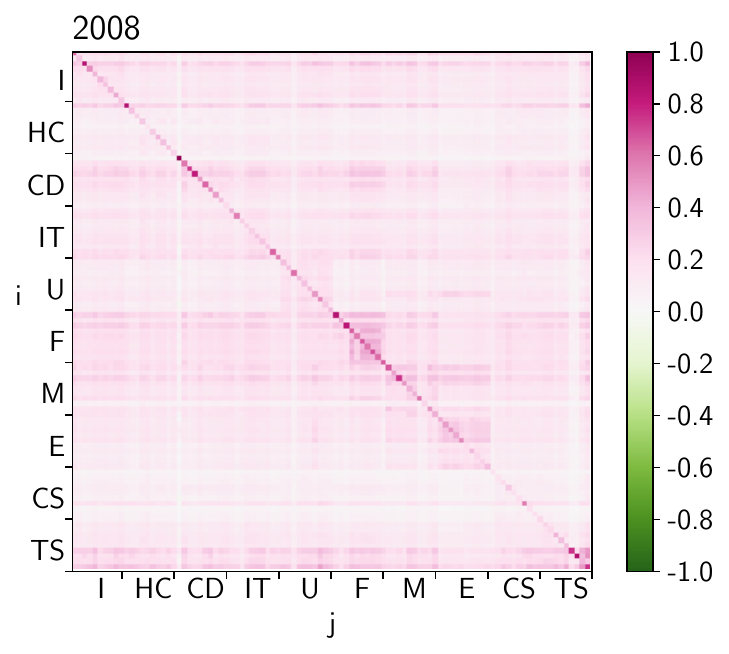}}%
	\end{minipage}
	\begin{minipage}{0.45\textwidth}
		{\includegraphics[width=1.\linewidth]{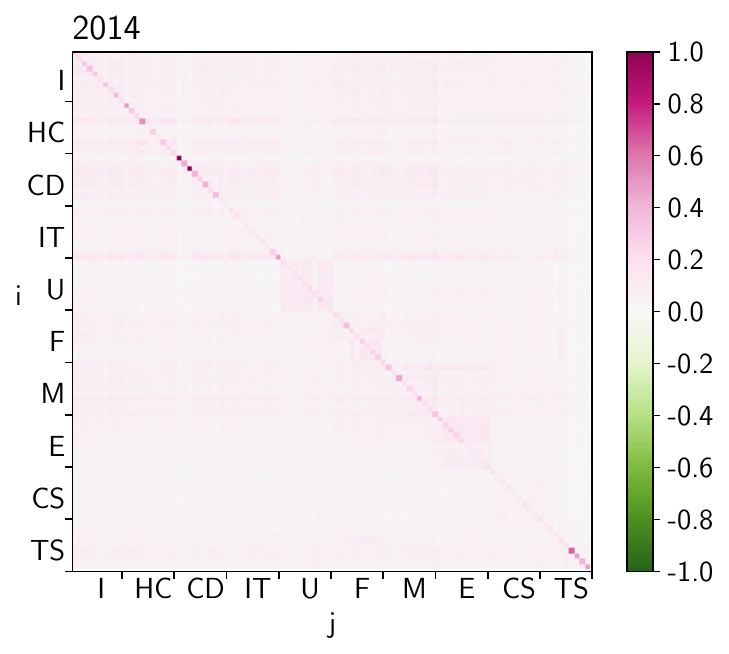}}%
	\end{minipage}%
	\begin{minipage}{0.45\textwidth}
		{\includegraphics[width=1.\linewidth]{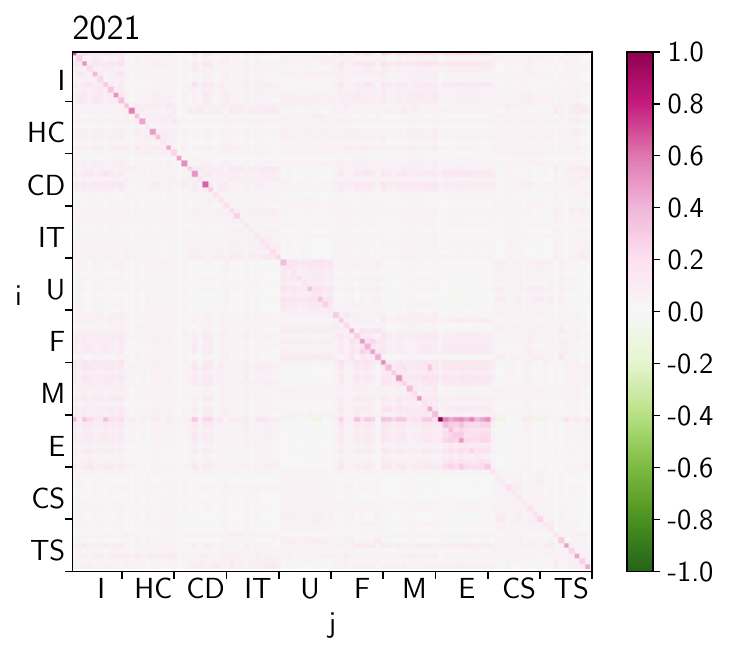}}%
	\end{minipage}
	\caption{\label{fig:ResponsesMat_2}Matrices for the normalized responses \textit{excluding} $\varepsilon_i(t) = 0$ with entries $\rho_{ij}(\tau)$ for $i,j=1,\ldots,99$ at the time lag $\tau=30$~s in 2007, 2008, 2014 and 2021. The stocks pairs $(i,j)$ correspond to the sectors industrials (I), health care (HC), consumer discretionary (CD), information technology (IT), utilities (U), financials (F), materials (M), energy (E), consumer staples (CS), and telecommunications services (TS).}
\end{figure}
For all four years and a time lag of $\tau=30\,$s, the corresponding matrices in Figs.~\ref{fig:ResponsesMat} and \ref{fig:ResponsesMat_2} show two kinds of gross patterns. First, as known from (symmetric) Pearson correlation matrices in financial markets~\cite{Laloux_1999,Plerou_2002,Guhr_Kaelber_2003}, there are blocks along the main diagonal corresponding to industrial sectors which are uncovered after sorting those matrices according to the ten industrial sectors, see~Sec.~\ref{sec:DataSet}. Second, there are vertical and horizontal stripes that sometimes extend across all industrial sectors.
For 2008, when we exclude $\varepsilon_j(t)=0$, the financial sector (F) has a more prominent block structure, while in 2021 the Energy sector (E) shows stronger responses when we include and exclude $\varepsilon_j(t)=0$. The vertical stripes emerge in 2007, 2008 and 2014 and almost not noticeable in 2021 when we include $\varepsilon_i(t) = 0$. Excluding the case  $\varepsilon_i(t) = 0$, those stripes are replaced by broader, horizontal stripes in 2008. 

\subsection{\label{sec:ActivePassiveResponseFunctions}Active and Passive Responses}

To examine the structures of the market cross--responses more closely,
we average for a fixed $i$ over all (pairwise) cross--responses
\begin{equation}
	\overline{R}_i^{(p)}(\tau)=\left\langle R_{ij}(\tau)\right\rangle_{j} \;,
	\label{eq33} \quad \text{with $j \neq i$}
\end{equation}
and for a fixed $j$ over all (pairwise) cross--responses
\begin{equation}
	\overline{R}_j^{(a)}(\tau)=\left\langle R_{ij}(\tau)\right\rangle_{i} \;,
	\label{eq35}  \quad \text{with $i \neq j$} \,.
\end{equation}
The former is referred to as passive response function, the latter as active response function~\cite{Wang2016_2,WangGuhr_2017}. The passive response function $\overline{R}_i^{(p)}(\tau)$ quantifies the influence of all trade signs in the market on the return of stock $i$ with time lag $\tau$, the active response function $\overline{R}_j^{(a)}(\tau)$ measures the influence of a trade or trades in stock $j$ on the returns of all other stocks with time lag $\tau$. In other words, index $i$ in $\overline{R}_i^{(p)}(\tau)$ labels an influenced or impacted stock and index $j$ in $\overline{R}_j^{(a)}(\tau)$ labels an influencing or impacting stock.
Of course, every stock is influenced and is influencing at the same time.

\begin{figure}[htbp]
	\captionsetup[subfigure]{labelformat=empty}
	\centering
	\begin{minipage}{0.5\textwidth}
	\subfloat[]{\begin{overpic}[width=1.0\linewidth]{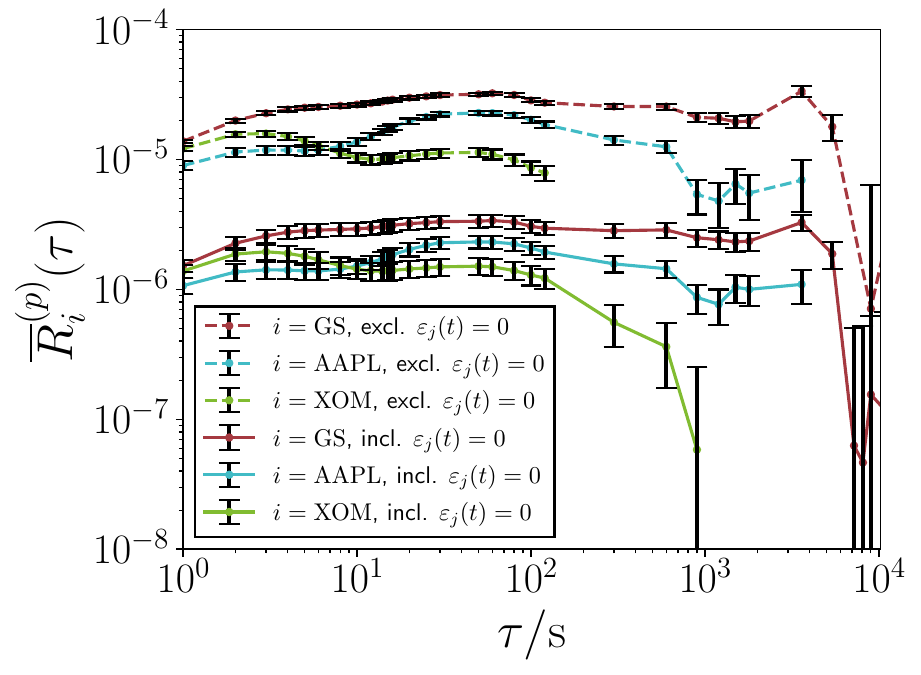}
			\put(61,22){\noindent\fbox{\parbox{0.8cm}{\sffamily 2007}}}
	\end{overpic}}%
\end{minipage}%
\begin{minipage}{0.5\textwidth}
	\subfloat[]{\begin{overpic}[width=1.0\linewidth]{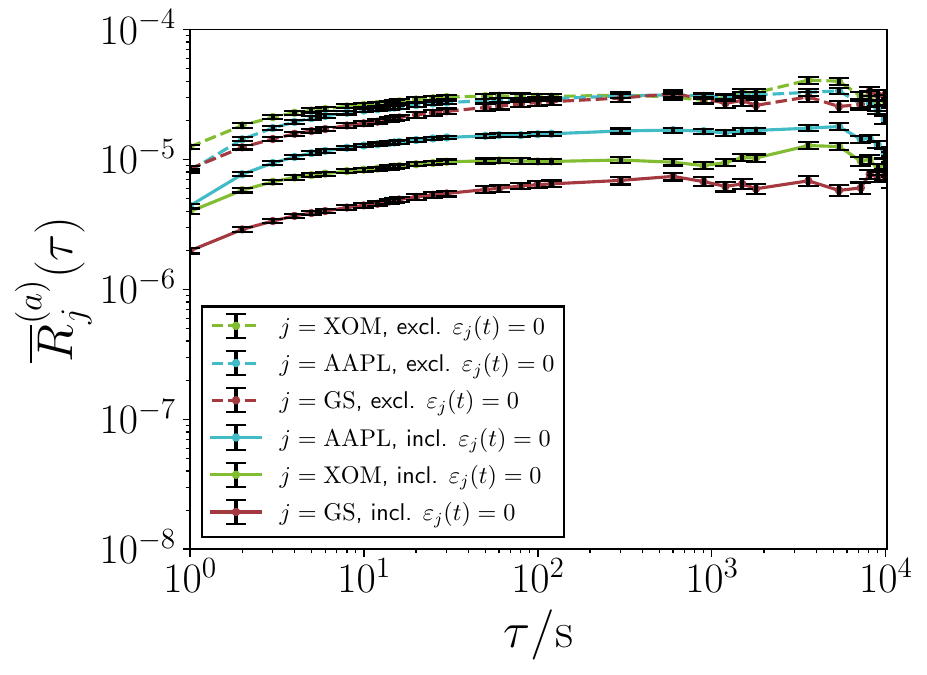}
			\put(71,22){\noindent\fbox{\parbox{0.8cm}{\sffamily 2007}}}
	\end{overpic}}%
\end{minipage}
	\begin{minipage}{0.5\textwidth}
	\subfloat[]{\begin{overpic}[width=1.0\linewidth]{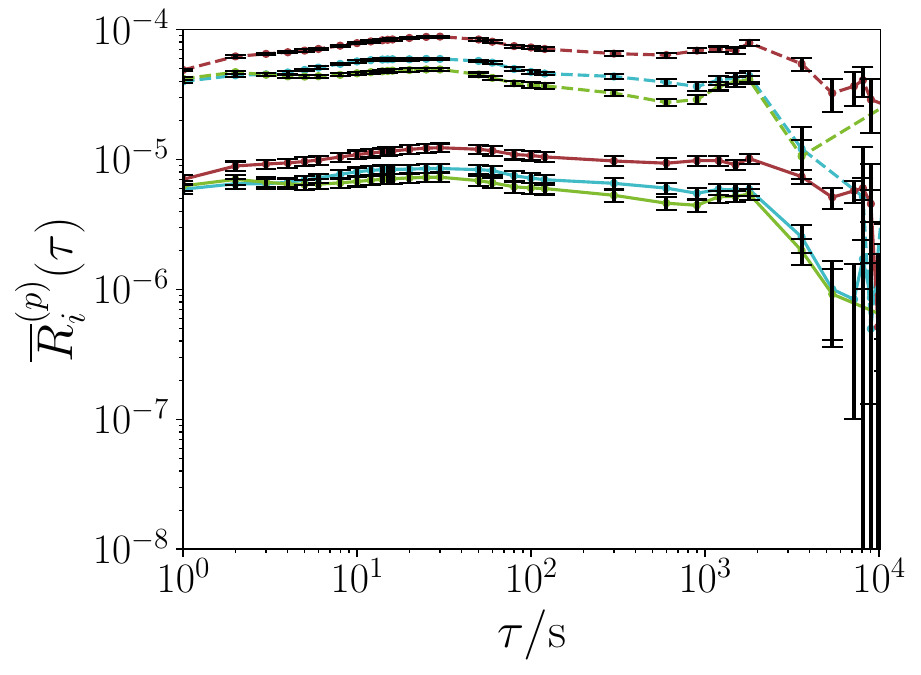}
			\put(25,25){\noindent\fbox{\parbox{0.8cm}{\sffamily 2008}}}
	\end{overpic}}%
\end{minipage}%
\begin{minipage}{0.5\textwidth}
	\subfloat[]{\begin{overpic}[width=1.0\linewidth]{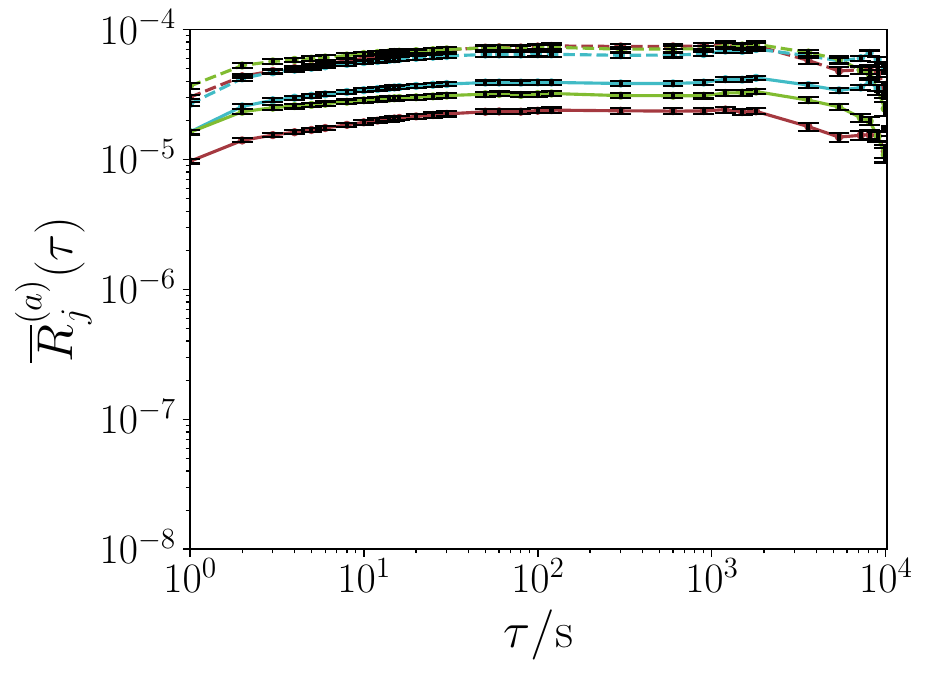}
			\put(25,25){\noindent\fbox{\parbox{0.8cm}{\sffamily 2008}}}
	\end{overpic}}%
\end{minipage}
	\caption{\label{fig:MarketResponse_IndStocks_20072008}Passive and active cross-response functions $\overline{R}_i^{(p)}(\tau)$ and $\overline{R}_j^{(a)}(\tau)$ for $i,j=\textrm{AAPL},\textrm{GS},\textrm{XOM}$ in the years 2007 and 2008 versus time lag $\tau$. Negative values for passive and active cross--responses are removed.}
\end{figure}

\begin{figure}[htbp]
	\captionsetup[subfigure]{labelformat=empty}
	\centering
	\begin{minipage}{0.5\textwidth}
		\subfloat[]{\begin{overpic}[width=1.0\linewidth]{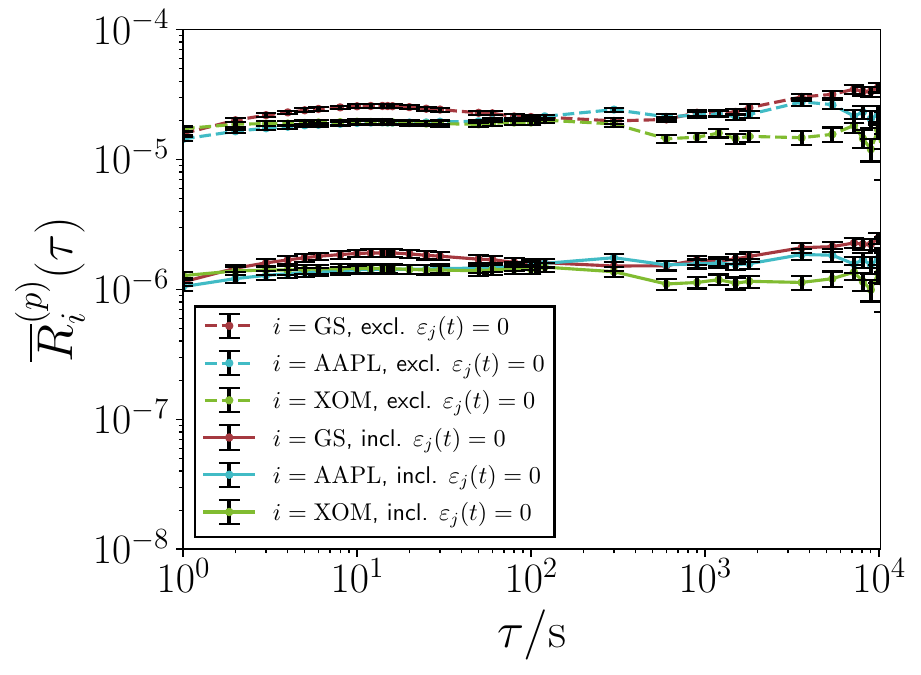}
				\put(70,25){\noindent\fbox{\parbox{0.8cm}{\sffamily 2014}}}
		\end{overpic}}%
	\end{minipage}%
	\begin{minipage}{0.5\textwidth}
		\subfloat[]{\begin{overpic}[width=1.0\linewidth]{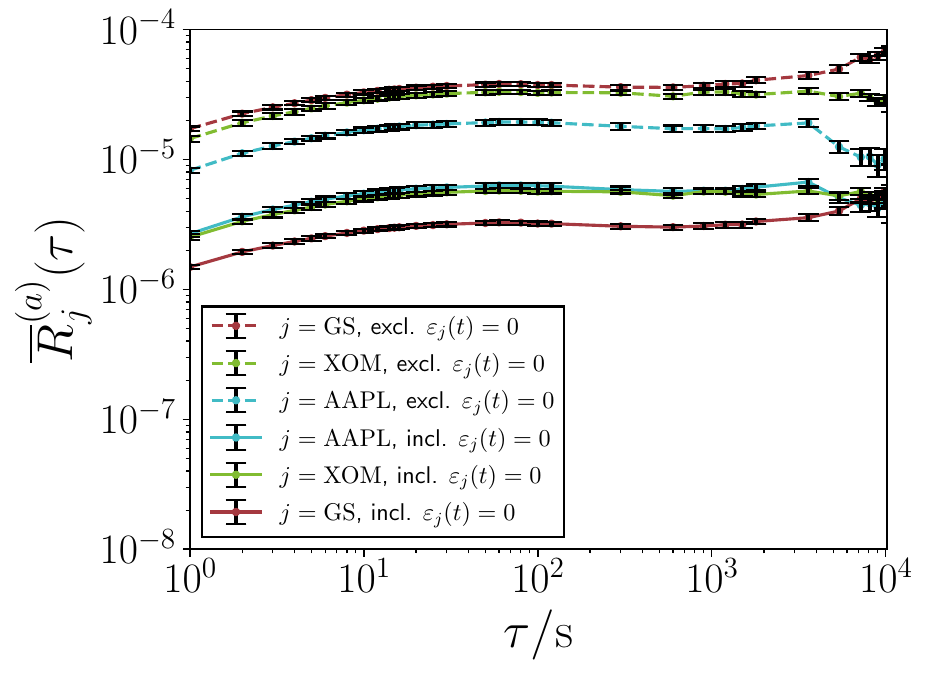}
				\put(70,25){\noindent\fbox{\parbox{0.8cm}{\sffamily 2014}}}
		\end{overpic}}%
	\end{minipage}
	\begin{minipage}{0.5\textwidth}
		\subfloat[]{\begin{overpic}[width=1.0\linewidth]{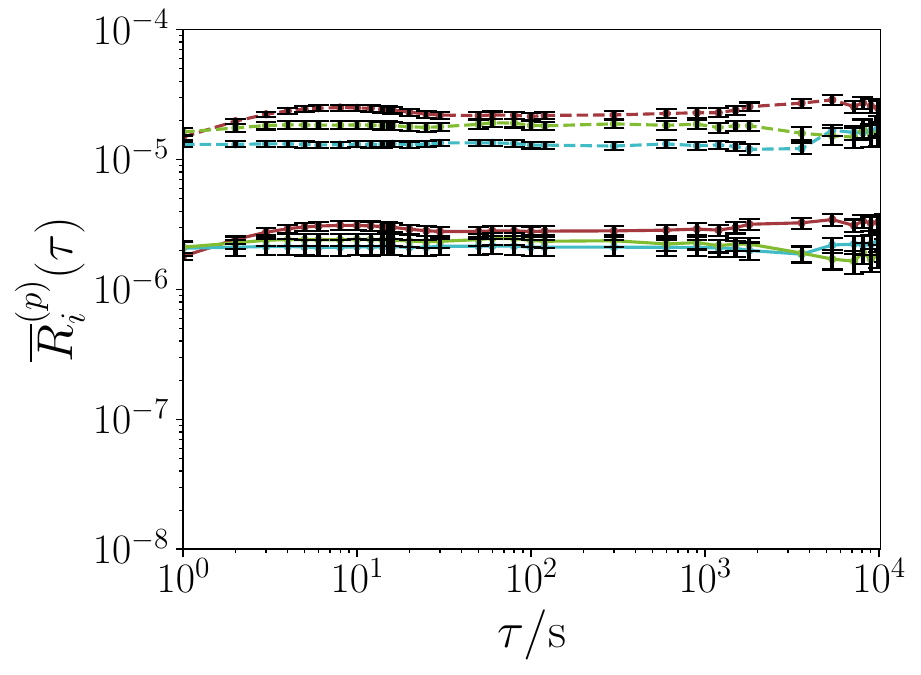}
				\put(25,25){\noindent\fbox{\parbox{0.8cm}{\sffamily 2021}}}
		\end{overpic}}%
	\end{minipage}%
	\begin{minipage}{0.5\textwidth}
		\subfloat[]{\begin{overpic}[width=1.0\linewidth]{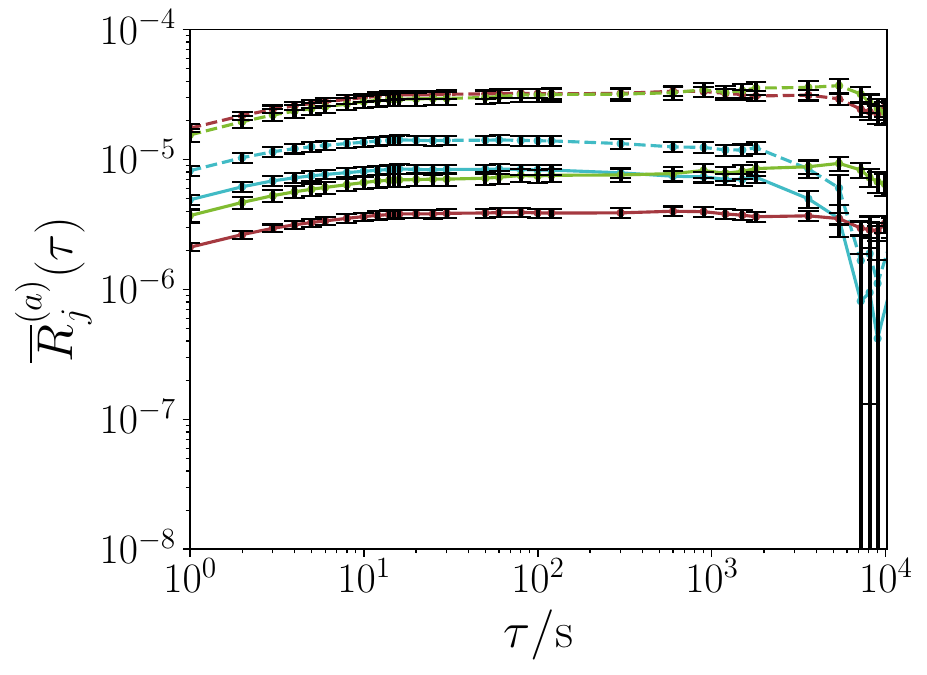}
				\put(25,25){\noindent\fbox{\parbox{0.8cm}{\sffamily 2021}}}
		\end{overpic}}%
	\end{minipage}
	\caption{\label{fig:MarketResponse_IndStocks_20142021}Passive and active cross-response functions $\overline{R}_i^{(p)}(\tau)$ and $\overline{R}_j^{(a)}(\tau)$ for $i,j=\textrm{AAPL},\textrm{GS},\textrm{XOM}$ in the years 2014 and 2021 versus time lag $\tau$. Negative values for passive and active cross--responses are removed.}
\end{figure}

Remarkably, Figs.~\ref{fig:MarketResponse_IndStocks_20072008} and \ref{fig:MarketResponse_IndStocks_20142021} show a qualitative different behavior for active and passive responses.
In Refs.~\cite{WSG2015preprint,Wang2016_2}, all stocks of the S\&P~500 index in the respective year were averaged. Here, we exclusively use the stocks sorted according to the GICS, see~Sec.~\ref{sec:DataSet}. The results are qualitatively similar.
In 2007 and 2008, the passive responses for AAPL and XOM begin to decrease after a much smaller time lag of length between $1000$ and $10000$ seconds.
For 2014 and 2021, there is no separation of two different time scales between passive and active response. As seen for the market self-- and cross--responses in Sec.~\ref{sec:MarketResponseFunctions}, neither shows a systemic downward trend except for the active cross--response in 2021 for \textsc{AAPL}.
The passive and active responses including $\varepsilon_i(t) = 0$ are always smaller than those excluding $\varepsilon_i(t) = 0$.

\section{\label{sec:TMA}Theoretical and Modeling Aspects}

In the pioneering works on response functions~\cite{Bouchaud_2003} and
price impacts \cite{LilloFarmer+2004}, two at first sight competing,
yet largely consistent, explanations for the non--Markovian effects
were offered.
In Ref.~\cite{Bouchaud_2003}, the self–responses (response of the price to
trading of this stock) were introduced and successfully used to uncover
non–Markovian effects on intraday timescales. In Ref.~\cite{LilloFarmer+2004}, long–term memory effects in order flow were
analyzed by studying autocorrelations of order signs. In line with this,
the authors investigated how liquidity fluctuations function to maintain
market efficiency.

Both explanations were formulated in mathematical models.
Ref.~\cite{Bouchaud_2003} introduces the transient price
impact model and provides explicit formulae for the
self--response whose key ingredient is a ``bare'' self--impact
function depending on the time lag $\tau$. The shape of this function
encodes how the non--Markovian effects act, in
Ref.~\cite{Bouchaud_2003} an algebraic form is assumed, corresponding
to a powerlaw dependence on the time lag. We mention in passing and to
avoid confusion in the terminology that this bare self--impact
function formally plays a role comparable to the ``response function''
in response theory as applied in traditional statistical mechanics and condensed matter theory. Here,
we use the term response function in a related, but different way as
defined in the cited econophysics literature. The qualitative
introduction and discussion of the permanent price impact model in
Ref.~\cite{LilloFarmer+2004} was complemented by a
mathematical model compared to the transient price impact model~\cite{BOUCHAUD_2009}. A related model for
order splitting was already put forward in Ref.~\cite{Lillo_2005}.

All these studies were confined to the self--responses or
self--impact, respectively. The early studies in the economics
literature on various cross--effects~\cite{Chordia_2000, HASBROUCK_2001, Pasquariello_2013, Boulatov_2012} also contain modeling aspects. The
first empirical results on the cross--response functions and the
trade--sign cross--correlations~\cite{WSG2015preprint,Wang2016,Wang2016_2} prompted theoretical
work and extensions of the above models. Simultaneously but
independently, such studies were put forward in Refs.~\cite{Benzaquen_2017}
and~\cite{WangGuhr_2017}. For the cross--response functions they involve a
``bare'' cross--impact function depending on the time lag
$\tau$. While the bare self--impact is driven by non--Markovian
liquidity effects, the bare cross--impact function must be entirely
information--driven. Detailed empirical analyses were also carried out
in Refs.~\cite{Benzaquen_2017} and~\cite{WangGuhr_2017}.  Various details in the fine
structure of cross--responses and trade--sign cross--correlations were
subsequently investigated. In Ref.~\cite{WNG2018}, the market
collective responses were statistically analyzed employing singular
value decomposition. The crucial role of the traded volume was
revealed which in turn is related to liquidity changes and
imbalances. In Ref.~\cite{WNG2019}, asymmetry in the information
affecting market impacts was analyzed and the model of
Ref.~\cite{WangGuhr_2017} was extended accordingly.

We also mention briefly that there has been substantial effort in extending theoretical frameworks from thermodynamics and dynamical systems to incorporate non--equilibrium dissipation and non--stationarity \cite{MARCONI2008111,Jepps_2016}. This has led, for example, to generalized fluctuation relations and the introduction of dissipation functions as theoretical tools to describe the evolution of systems in traditional physics. While these developments provide valuable conceptual background, our approach here does not build on such general formulations.
This would be rewarding but a formidable task, as the fundamental equations describing the dynamics of financial markets are not known, and the underlying mechanisms themselves change over time.
Instead, we focus on data analysis and the established response function methodology of econophysics. In a statistical setting for complex systems, the approach of \cite{WangGuhr_2017} provides information on the dynamical evolution of traders’ behavior and their interactions in a single stock as well as across different stocks. Both short--term price changes, resulting from temporary liquidity shortages, and long--term price changes, which may be driven by private information, significantly influence trading behavior.

\section{\label{Sec:Discussion}Conclusions}

We considerably extended the studies of
Refs.~\cite{WSG2015preprint,Wang2016,Wang2016_2} to investigate the
non--stationarity of non--Markovian effects.  We found in the dynamics
of the trading sizable changes in the dynamics which reveals severe
modifications in the traders' actions.  The market self-- and
cross--responses and the market trade sign self-- and
cross--correlators facilitate, at least their averaged versions,
rather general observations and statements.  How do the different
results fit together? To what extent is the crisis year 2008 special?

There are two trends to be recognized over the years.
First, the responses for the crisis year of 2008 are larger than for the other years. The self--responses for 2014 and 2021 remain on a higher level compared to 2007.
Second, both market self--responses and cross--responses are strikingly less concave. The self--responses of 2014 and 2021 indicate that self--responses increasing for larger time lags are quite stable over time.
Larger response values in times of crises and responses increasing for larger time lags may have different causes.

Around and during the financial crisis of 2008/2009~\cite{Heckens_2020,heckens2021new,heckens2022new}, algorithmic trading also increased dramatically~\cite{kissell2020algorithmic}. Over the time period covered by our study, algorithmic trading has almost completely replaced traditional trading. Thus, trading may have been affected by both phenomena, the financial crisis itself as well as the surge in algorithmic trading.
It is important to note that algorithmic trading does not only include high--frequency trades on very short time scales.

The increase in the response function may be related to the traders' reaction to the financial crisis itself, as it increases from 2007 to 2008 within a relatively short period of time.
In the liquidity game, liquidity providers are more alert placing their orders at more expensive price levels for the liquidity takers, \textit{i.e.} liquidity providers demand a larger risk premium for compensation. The higher values in the self--responses for 2014 and 2021 compared to 2007 indicate that the crisis has had a lasting influence on the trader's actions.
Nevertheless, the cross--responses drop back to magnitudes similar to 2007.
The shape of the self-- and cross--responses is no longer concave for 2014 and 2021 hinting at different causes, \textit{i.e.} the algorithmic trading might have changed the responses. %
One of the most notable changes in the traders' behavior around 2008 is the change in passive and active cross--responses and might also be related to the change in algorithmic trading.
The time scales of the passive and active response functions are different for 2007 and 2008, while we do not observe such behavior for 2014 and 2021.

Another interpretation is also possible if we assume that 
the change in response functions and the emergence of algorithmic trading coincide by chance. We know that the market operates in a new mode in terms of self--responses increasing for larger time lags. Even in a trading environment with exclusively human or algorithmic traders the market could have made a transition into a new stable state of trading due to a crisis that destabilized the subtle balance of the traders' decisions. 

The market trade sign correlators provide an indication of a major change from 2007 to 2008, moving towards a shorter memory
between different stocks. A shorter memory in the liquidity game means that liquidity takers try to carry out their trading strategies more quickly.
For 2014, the memory for the trade sign correlators becomes more similar to those of 2007. For 2021, the memory decreases again which might be explained by faster algorithmic trading.
It is worth noting that the trade sign self--correlators are clearly different in 2021, with a second regime emerging. This regime is less noticeable for the other years. 
The second regime hints at a longer memory in the traders' actions for larger time lags which might be related to a different traders' strategy. The diffusive behavior of the returns resulting from a balance between sub-- and superdiffusive dynamics behavior strongly depends on the time lag $\tau$.

In short, trading changed its phenomenology. The reasons cannot be fully disentangled, but important factors for the change in the traders' interactions may be increased risk aversion due to the financial crisis of 2008/2009 and the rise of algorithmic trading.
Our results indicate that the market mechanisms as borne out in the details of the order book are remarkably non--stationary. While the rules of the game do not change, it is now played in a different way. Close monitoring and careful data analysis over the years to come seems called for.

\section*{Acknowledgment}

We thank Shanshan Wang for fruitful discussions, we are particularly
grateful to R\"udiger Kiesel for helpful remarks on the economics
literature.

\clearpage

\bibliography{Lit.bib}

\clearpage

\pagenumbering{gobble}

\begin{appendices}

\section{\label{sec:Tables}{Industrial Sectors and Stocks}}

\begin{table*}
	\linespread{0.5}
	\caption{99 stocks from ten economic sectors for 2008.} 
	\begin{center}
		\begin{footnotesize}
			\begin{tabular}{llcll} 
				\toprule
				\multicolumn{2}{l}{\textbf{Industrials} (I)} &~~& \multicolumn{2}{l}{\textbf{Financials} (F)}  \vspace{0.1cm} \\
				Symbol	&Company				 	&	&Symbol		&Company						\\
				\midrule
				FLR		&Fluor Corp. (New)				&	&CME		&CME Group Inc.					\\
				LMT		&Lockheed Martin Corp.			&	&GS			&Goldman Sachs Group				\\
				FLS		&Flowserve Corporation			&	&ICE			&Intercontinental Exchange Inc.		\\
				PCP		&Precision Castparts			&	&AVB		&AvalonBay Communities				\\
				LLL		&L-3 Communications Holdings	&	&BEN		&Franklin Resources					\\
				UNP		&Union Pacific					&	&BXP		&Boston Properties				 	\\
				BNI		&Burlington Northern Santa Fe C &	&SPG		&Simon Property 	Group  Inc	 		\\
				FDX		&FedEx Corporation				&	&VNO		&Vornado Realty Trust				\\
				GWW	&Grainger (W.W.) Inc.			&	&PSA		&Public Storage					\\
				GD		&General Dynamics				&	&MTB		&M$\&$T Bank Corp.				\\
				\midrule
				\multicolumn{2}{l}{\textbf{Health Care} (HC)} &~& \multicolumn{2}{l}{\textbf{Materials} (M)} 	\vspace{0.1cm}  \\
				Symbol	&Company					&	&Symbol		&Company						\\
				\midrule
				ISRG	&Intuitive Surgical Inc.			&	 &X			&United States Steel Corp.			\\
				BCR		&Bard (C.R.) Inc.				&	 &MON		&Monsanto Co.						\\
				BDX		&Becton  Dickinson				&	 &CF			&CF Industries Holdings Inc			\\
				GENZ	&Genzyme Corp.				&	 &FCX		&Freeport-McMoran Cp $\&$ Gld 		\\
				JNJ		&Johnson $\&$ Johnson			&	&APD		&Air Products $\&$ Chemicals			\\
				LH		&Laboratory Corp. of America Holding & &PX		&Praxair  Inc.						\\
				ESRX	&Express Scripts				&	&VMC		&Vulcan Materials					\\
				CELG	&Celgene Corp.				&	&ROH		&Rohm $\&$ Haas					\\
				ZMH		&Zimmer Holdings				&	&NUE		&Nucor Corp.						\\
				AMGN	&Amgen						&	&PPG		&PPG Industries					\\
				\midrule
				\multicolumn{2}{l}{\textbf{Consumer	Discretionary} (CD)} &~& \multicolumn{2}{l}{\textbf{Energy} (E)}  \vspace{0.1cm}  \\
				Symbol	&Company					&	&Symbol		&Company					\\
				\midrule
				WPO	&Washington Post			&	&RIG		&Transocean Inc. (New)				\\
				AZO		&AutoZone Inc.					&	&APA		&Apache Corp.					\\
				SHLD	&Sears Holdings Corporation		&	&EOG		&EOG Resources					\\
				WYNN	&Wynn Resorts Ltd.				&	&DVN		&Devon Energy Corp.				\\
				AMZN	&Amazon Corp.				&	&HES		&Hess Corporation					\\
				WHR	&Whirlpool Corp.			&	&XOM		&Exxon Mobil Corp.					\\
				VFC		&V.F. Corp.					&	&SLB		&Schlumberger Ltd.					\\
				APOL	&Apollo Group					&	&CVX		&Chevron Corp.					\\
				NKE		&NIKE Inc.					&	&COP		&ConocoPhillips					\\
				MCD		&McDonald's Corp.				&	&OXY		&Occidental Petroleum				\\
				\midrule
				\multicolumn{2}{l}{\textbf{Information Technology} (IT)} &~& \multicolumn{2}{l}{\textbf{Consumer Staples} (CS)}  \vspace{0.1cm} \\
				Symbol	&Company					&	&Symbol		&Company						\\
				\midrule
				GOOG	&Google Inc.					&	&BUD		&Anheuser-Busch					\\
				MA		&Mastercard Inc.			&	&PG			&Procter $\&$ Gamble				\\
				AAPL	&Apple Inc.					&	&CL			&Colgate-Palmolive					\\
				IBM		&International Bus. Machines		&	&COST		&Costco Co.						\\
				MSFT	&Microsoft Corp.			& 	&WMT		&Wal-Mart Stores					\\
				CSCO	&Cisco Systems				&	&PEP		&PepsiCo Inc.						\\
				INTC		&Intel Corp.					&	&LO			&Lorillard Inc.						\\
				QCOM	&QUALCOMM Inc.				&	&UST		&UST Inc.							\\
				CRM		&Salesforce Com Inc. 			&	&GIS		&General Mills						\\
				WFR		&MEMC Electronic Materials		&	&KMB		&Kimberly-Clark					\\
				\midrule
				\multicolumn{2}{l}{\textbf{Utilities} (U)} &~& \multicolumn{2}{l}{\textbf{Telecommunications Services} (TS)} \vspace{0.1cm}  \\
				Symbol	&Company					&	&Symbol		&Company						\\
				\midrule
				ETR		&Entergy Corp.					&	&T			&AT$\&$T Inc.						\\
				EXC		&Exelon Corp.					&	&VZ			&Verizon Communications			\\
				CEG		&Constellation Energy  Group 		&	&EQ			&Embarq Corporation				\\
				FE		&FirstEnergy Corp.				&	&AMT		&American Tower Corp.				\\
				FPL		&FPL Group					&	&CTL 		&Century Telephone					\\
				SRE		&Sempra Energy				&	&S			&Sprint Nextel Corp.					\\
				STR		&Questar Corp.					&	&Q			&Qwest Communications  Int 			\\
				TEG		&Integrys Energy Group Inc. 		&	&WIN		&Windstream Corporation			\\
				EIX		&Edison Int'l					&	&FTR		&Frontier Communications			\\
				AYE		&Allegheny Energy				&	&			&									\\
				\bottomrule
			\end{tabular}
		\end{footnotesize}
	\end{center}
	\label{tab:stocks_2008}
\end{table*}

\begin{table*}
	\linespread{0.5}
	\caption{99 stocks from ten economic sectors for 2007. Stocks in bold represent new stocks compared to 2008.} 
	\begin{center}
		\begin{footnotesize}
			\begin{tabular}{llcll} 
				\toprule
				\multicolumn{2}{l}{\textbf{Industrials} (I)} &~~& \multicolumn{2}{l}{\textbf{Financials} (F)}  \vspace{0.1cm} \\
				Symbol	&Company				 	&	&Symbol		&Company						\\
				\midrule
				FLR		&Fluor Corp. (New)				&	&CME		&CME Group Inc.					\\
				LMT		&Lockheed Martin Corp.			&	&GS			&Goldman Sachs Group				\\
				FLS		&Flowserve Corporation			&	&ICE			&Intercontinental Exchange Inc.		\\
				PCP		&Precision Castparts			&	&AVB		&AvalonBay Communities				\\
				LLL		&L-3 Communications Holdings	&	&BEN		&Franklin Resources					\\
				UNP		&Union Pacific					&	&BXP		&Boston Properties				 	\\
				\textbf{MMM}		& \textbf{3M Company} &	&SPG		&Simon Property 	Group  Inc	 		\\
				FDX		&FedEx Corporation				&	&VNO		&Vornado Realty Trust				\\
				GWW	&Grainger (W.W.) Inc.			&	&PSA		&Public Storage					\\
				GD		&General Dynamics				&	&MTB		&M$\&$T Bank Corp.				\\
				\midrule
				\multicolumn{2}{l}{\textbf{Health Care} (HC)} &~& \multicolumn{2}{l}{\textbf{Materials} (M)} 	\vspace{0.1cm}  \\
				Symbol	&Company					&	&Symbol		&Company						\\
				\midrule
				ISRG	&Intuitive Surgical Inc.			&	 &X			&United States Steel Corp.			\\
				BCR		&Bard (C.R.) Inc.				&	 &MON		&Monsanto Co.						\\
				BDX		&Becton  Dickinson				&	 &CF			&CF Industries Holdings Inc			\\
				\textbf{ABT}	& \textbf{Abbott Laboratories}				&	 &FCX		&Freeport-McMoran Cp $\&$ Gld 		\\
				JNJ		&Johnson $\&$ Johnson			&	&APD		&Air Products $\&$ Chemicals			\\
				LH		&Laboratory Corp. of America Holding & &PX		&Praxair  Inc.						\\
				ESRX	&Express Scripts				&	&VMC		&Vulcan Materials					\\
				CELG	&Celgene Corp.				&	&\textbf{SIAL}		&\textbf{Sigma Aldrich Corp.}					\\
				ZMH		&Zimmer Holdings				&	&NUE		&Nucor Corp.						\\
				AMGN	&Amgen						&	&PPG		&PPG Industries					\\
					&						&	& \textbf{SHW}		&\textbf{Sherwin--Williams Company} \\
				\midrule
				\multicolumn{2}{l}{\textbf{Consumer	Discretionary} (CD)} &~& \multicolumn{2}{l}{\textbf{Energy} (E)}  \vspace{0.1cm}  \\
				Symbol	&Company					&	&Symbol		&Company					\\
				\midrule
				AZO	& 	AutoZone Inc.	&	&\textbf{NBL}		&\textbf{Noble Energy Inc.}			\\
				SHLD		&Sears Holdings Corporation					&	&APA		&Apache Corp.					\\
				WYNN	&Wynn Resorts Ltd.		&	&EOG		&EOG Resources					\\
				AMZN	&Amazon Corp.				&	&DVN		&Devon Energy Corp.				\\
				WHR	&Whirlpool Corp.				&	&HES		&Hess Corporation					\\
				VFC	&V.F. Corp.			&	&XOM		&Exxon Mobil Corp.					\\
				APOL	&Apollo Group					&	&SLB		&Schlumberger Ltd.					\\
				NKE	&NIKE Inc.					&	&CVX		&Chevron Corp.					\\
				MCD			&McDonald's Corp.					&	&COP		&ConocoPhillips					\\
						&				&	&OXY		&Occidental Petroleum				\\
				\midrule
				\multicolumn{2}{l}{\textbf{Information Technology} (IT)} &~& \multicolumn{2}{l}{\textbf{Consumer Staples} (CS)}  \vspace{0.1cm} \\
				Symbol	&Company					&	&Symbol		&Company						\\
				\midrule
				GOOG	&Google Inc.					&	&\textbf{CLX}		&\textbf{The Colrox Company}					\\
				MA		&Mastercard Inc.			&	&PG			&Procter $\&$ Gamble				\\
				AAPL	&Apple Inc.					&	&CL			&Colgate-Palmolive					\\
				IBM		&International Bus. Machines		&	&COST		&Costco Co.						\\
				MSFT	&Microsoft Corp.			& 	&WMT		&Wal-Mart Stores					\\
				CSCO	&Cisco Systems				&	&PEP		&PepsiCo Inc.						\\
				INTC		&Intel Corp.					&	&\textbf{KO}			&\textbf{The Coca-Cola Company}						\\
				QCOM	&QUALCOMM Inc.				&	&\textbf{K}		&\textbf{Kellogg Company}							\\
				CRM		&Salesforce Com Inc. 			&	&GIS		&General Mills						\\
				WFR		&MEMC Electronic Materials		&	&KMB		&Kimberly-Clark					\\
				\midrule
				\multicolumn{2}{l}{\textbf{Utilities} (U)} &~& \multicolumn{2}{l}{\textbf{Telecommunications Services} (TS)} \vspace{0.1cm}  \\
				Symbol	&Company					&	&Symbol		&Company						\\
				\midrule
				ETR		&Entergy Corp.					&	&T			&AT$\&$T Inc.						\\
				EXC		&Exelon Corp.					&	&VZ			&Verizon Communications			\\
				CEG		&Constellation Energy  Group 		&	&\textbf{DTV}			&\textbf{DirecTV Group Inc.}				\\
				FE		&FirstEnergy Corp.				&	&AMT		&American Tower Corp.				\\
				\textbf{PCG}		&\textbf{PG$\&E$ Corp}					&	&CTL 		&Century Telephone					\\
				SRE		&Sempra Energy				&	&S			&Sprint Nextel Corp.					\\
				\textbf{D}		&\textbf{Dominion Energy Inc.}					&	&\textbf{TLAB}			&\textbf{Tellabs Inc.} 			\\
				TEG		&Integrys Energy Group Inc. 		&	&WIN		&Windstream Corporation			\\
				EIX		&Edison Int'l					&	&\textbf{CMCSA}		&\textbf{Comcast Corp.}			\\
				\textbf{SO}		&\textbf{The Southern Company}				&	&			&									\\
				\bottomrule
			\end{tabular}
		\end{footnotesize}
	\end{center}
	\label{tab:stocks_2007}
\end{table*}

\begin{table*}
	\linespread{0.5}
	\caption{99 stocks from ten economic sectors for 2014. Stocks in bold represent new stocks compared to 2008.} 
	\begin{center}
		\begin{footnotesize}
			\begin{tabular}{llcll} 
				\toprule
				\multicolumn{2}{l}{\textbf{Industrials} (I)} &~~& \multicolumn{2}{l}{\textbf{Financials} (F)}  \vspace{0.1cm} \\
				Symbol	&Company				 	&	&Symbol		&Company						\\
				\midrule
				FLR		&Fluor Corp. (New)				&	&CME		&CME Group Inc.					\\
				LMT		&Lockheed Martin Corp.			&	&GS			&Goldman Sachs Group				\\
				FLS		&Flowserve Corporation			&	&ICE			&Intercontinental Exchange Inc.		\\
				PCP		&Precision Castparts			&	&AVB		&AvalonBay Communities				\\
				LLL		&L-3 Communications Holdings	&	&BEN		&Franklin Resources					\\
				UNP		&Union Pacific					&	&BXP		&Boston Properties				 	\\
				\textbf{AAL}		&\textbf{American Airlines Group Inc.} &	&SPG		&Simon Property 	Group  Inc	 		\\
				FDX		&FedEx Corporation				&	&VNO		&Vornado Realty Trust				\\
				GWW	&Grainger (W.W.) Inc.			&	&PSA		&Public Storage					\\
				GD		&General Dynamics				&	&MTB		&M$\&$T Bank Corp.				\\
				\midrule
				\multicolumn{2}{l}{\textbf{Health Care} (HC)} &~& \multicolumn{2}{l}{\textbf{Materials} (M)} 	\vspace{0.1cm}  \\
				Symbol	&Company					&	&Symbol		&Company						\\
				\midrule
				ISRG	&Intuitive Surgical Inc.			&	 &X			&United States Steel Corp.			\\
				BCR		&Bard (C.R.) Inc.				&	 &MON		&Monsanto Co.						\\
				BDX		&Becton  Dickinson				&	 &CF			&CF Industries Holdings Inc			\\
				\textbf{BIIB}	&\textbf{Biogen Idec Inc.}.				&	 &FCX		&Freeport-McMoran Cp $\&$ Gld 		\\
				JNJ		&Johnson $\&$ Johnson			&	&APD		&Air Products $\&$ Chemicals			\\
				LH		&Laboratory Corp. of America Holding & &PX		&Praxair  Inc.						\\
				ESRX	&Express Scripts				&	&VMC		&Vulcan Materials					\\
				CELG	&Celgene Corp.				&	&\textbf{DOW}		&\textbf{DOW Chemicals Co. Com.} \\
				ZMH		&Zimmer Holdings				&	&NUE		&Nucor Corp.						\\
				AMGN	&Amgen						&	&PPG		&PPG Industries					\\
				\midrule
				\multicolumn{2}{l}{\textbf{Consumer	Discretionary} (CD)} &~& \multicolumn{2}{l}{\textbf{Energy} (E)}  \vspace{0.1cm}  \\
				Symbol	&Company					&	&Symbol		&Company					\\
				\midrule
				\textbf{GHC}	&\textbf{Graham Holdings}			&	&RIG		&Transocean Inc. (New)				\\
				AZO		&AutoZone Inc.					&	&APA		&Apache Corp.					\\
				SHLD	&Sears Holdings Corporation		&	&EOG		&EOG Resources					\\
				WYNN	&Wynn Resorts Ltd.				&	&DVN		&Devon Energy Corp.				\\
				AMZN	&Amazon Corp.				&	&HES		&Hess Corporation					\\
				WHR	&Whirlpool Corp.			&	&XOM		&Exxon Mobil Corp.					\\
				VFC		&V.F. Corp.					&	&SLB		&Schlumberger Ltd.					\\
				APOL	&Apollo Group					&	&CVX		&Chevron Corp.					\\
				NKE		&NIKE Inc.					&	&COP		&ConocoPhillips					\\
				MCD		&McDonald's Corp.				&	&OXY		&Occidental Petroleum				\\
				\midrule
				\multicolumn{2}{l}{\textbf{Information Technology} (IT)} &~& \multicolumn{2}{l}{\textbf{Consumer Staples} (CS)}  \vspace{0.1cm} \\
				Symbol	&Company					&	&Symbol		&Company						\\
				\midrule
				GOOG	&Google Inc.					&	&BUD		&Anheuser-Busch					\\
				MA		&Mastercard Inc.			&	&PG			&Procter $\&$ Gamble				\\
				AAPL	&Apple Inc.					&	&CL			&Colgate-Palmolive					\\
				IBM		&International Bus. Machines		&	&COST		&Costco Co.						\\
				MSFT	&Microsoft Corp.			& 	&WMT		&Wal-Mart Stores					\\
				CSCO	&Cisco Systems				&	&PEP		&PepsiCo Inc.						\\
				INTC		&Intel Corp.					&	&LO			&Lorillard Inc.						\\
				QCOM	&QUALCOMM Inc.				&	&\textbf{MO}		&\textbf{Altria Group Inc.}						\\
				CRM		&Salesforce Com Inc. 			&	&GIS		&General Mills						\\
				\textbf{SUNE}		&\textbf{Sunedison Inc. Com.}		&	&KMB		&Kimberly-Clark					\\
				\midrule
				\multicolumn{2}{l}{\textbf{Utilities} (U)} &~& \multicolumn{2}{l}{\textbf{Telecommunications Services} (TS)} \vspace{0.1cm}  \\
				Symbol	&Company					&	&Symbol		&Company						\\
				\midrule
				ETR		&Entergy Corp.					&	&T			&AT$\&$T Inc.						\\
				EXC		&Exelon Corp.					&	&VZ			&Verizon Communications			\\
				\textbf{DUK}		&\textbf{Duke Energy Corp. Com.} 		&	&\textbf{FB}			&\textbf{Facebook Inc. Com.}				\\
				FE		&FirstEnergy Corp.				&	&AMT		&American Tower Corp.				\\
				\textbf{NEE}		&\textbf{NextEra Energy Inc. Com.}					&	&CTL 		&Century Telephone					\\
				SRE		&Sempra Energy				&	&S			&Sprint Nextel Corp.					\\
				STR		&Questar Corp.					&	&Q			&Qwest Communications  Int 			\\
				TEG		&Integrys Energy Group Inc. 		&	&WIN		&Windstream Corporation			\\
				EIX		&Edison Int'l					&	&FTR		&Frontier Communications			\\
				\textbf{SO}		&\textbf{Southern Com.}				&	&			&									\\
				\bottomrule
			\end{tabular}
		\end{footnotesize}
	\end{center}
	\label{tab:stocks_2014}
\end{table*}

\begin{table*}
	\linespread{0.5}
	\caption{99 stocks from ten economic sectors for 2021. Stocks in bold represent new stocks compared to 2014. Stocks in italicized represent new stocks compared to 2008.} 
	\begin{center}
		\begin{footnotesize}
			\begin{tabular}{llcll} 
				\toprule
				\multicolumn{2}{l}{\textbf{Industrials} (I)} &~~& \multicolumn{2}{l}{\textbf{Financials} (F)}  \vspace{0.1cm} \\
				Symbol	&Company				 	&	&Symbol		&Company						\\
				\midrule
				FLR		&Fluor Corp. (New)				&	&CME		&CME Group Inc.					\\
				LMT		&Lockheed Martin Corp.			&	&GS			&Goldman Sachs Group				\\
				FLS		&Flowserve Corporation			&	&ICE			&Intercontinental Exchange Inc.		\\
				\textbf{BA}		&\textbf{The Boeing Company}			&	&AVB		&AvalonBay Communities				\\
				\textbf{LHX}		&\textbf{L3 Harris Technologies}	&	&BEN		&Franklin Resources					\\
				UNP		&Union Pacific					&	&BXP		&Boston Properties				 	\\
				\textit{AAL}		&\textit{American Airlines Group Inc.} &	&SPG		&Simon Property 	Group  Inc	 		\\
				FDX		&FedEx Corporation				&	&VNO		&Vornado Realty Trust				\\
				GWW	&Grainger (W.W.) Inc.			&	&PSA		&Public Storage					\\
				GD		&General Dynamics				&	&MTB		&M$\&$T Bank Corp.				\\
				\midrule
				\multicolumn{2}{l}{\textbf{Health Care} (HC)} &~& \multicolumn{2}{l}{\textbf{Materials} (M)} 	\vspace{0.1cm}  \\
				Symbol	&Company					&	&Symbol		&Company						\\
				\midrule
				ISRG	&Intuitive Surgical Inc.			&	 &X			&United States Steel Corp.			\\
				\textbf{MRNA}		&\textbf{Moderna Inc.}				&	 &\textbf{VALE}		&\textbf{Vale S.A.}						\\
				BDX		&Becton  Dickinson				&	 &CF			&CF Industries Holdings Inc			\\
				\textit{BIIB}	&\textit{Biogen Idec Inc.}.				&	 &FCX		&Freeport-McMoran Cp $\&$ Gld 		\\
				JNJ		&Johnson $\&$ Johnson			&	&APD		&Air Products $\&$ Chemicals			\\
				LH		&Laboratory Corp. of America Holding & &\textbf{LIN}		&\textbf{Linde plc.}						\\
				\textbf{CI}	&\textbf{Cigna Corp}				&	&VMC		&Vulcan Materials					\\
				\textbf{BMY}	&\textbf{Bristol-Meyers Squibb}				&	&\textit{DOW}		&\textit{DOW Chemicals Co. Com.} \\
				\textbf{ZBH}		&\textbf{Zimmer Biomet Holdings Inc.}				&	&NUE		&Nucor Corp.						\\
				AMGN	&Amgen						&	&PPG		&PPG Industries					\\
				\midrule
				\multicolumn{2}{l}{\textbf{Consumer	Discretionary} (CD)} &~& \multicolumn{2}{l}{\textbf{Energy} (E)}  \vspace{0.1cm}  \\
				Symbol	&Company					&	&Symbol		&Company					\\
				\midrule
				\textit{GHC}	&\textit{Graham Holdings}			&	&RIG		&Transocean Inc. (New)				\\
				AZO		&AutoZone Inc.					&	&APA		&Apache Corp.					\\
				\textbf{TSLA}	&\textbf{Tesla Inc}		&	&EOG		&EOG Resources					\\
				WYNN	&Wynn Resorts Ltd.				&	&DVN		&Devon Energy Corp.				\\
				AMZN	&Amazon Corp.				&	&HES		&Hess Corporation					\\
				WHR	&Whirlpool Corp.			&	&XOM		&Exxon Mobil Corp.					\\
				VFC		&V.F. Corp.					&	&SLB		&Schlumberger Ltd.					\\
				\textbf{BABA}	&\textbf{Alibaba}					&	&CVX		&Chevron Corp.					\\
				NKE		&NIKE Inc.					&	&COP		&ConocoPhillips					\\
				MCD		&McDonald's Corp.				&	&OXY		&Occidental Petroleum				\\
				\midrule
				\multicolumn{2}{l}{\textbf{Information Technology} (IT)} &~& \multicolumn{2}{l}{\textbf{Consumer Staples} (CS)}  \vspace{0.1cm} \\
				Symbol	&Company					&	&Symbol		&Company						\\
				\midrule
				GOOG	&Google Inc.					&	&BUD		&Anheuser-Busch					\\
				MA		&Mastercard Inc.			&	&PG			&Procter $\&$ Gamble				\\
				AAPL	&Apple Inc.					&	&CL			&Colgate-Palmolive					\\
				IBM		&International Bus. Machines		&	&COST		&Costco Co.						\\
				MSFT	&Microsoft Corp.			& 	&WMT		&Wal-Mart Stores					\\
				CSCO	&Cisco Systems				&	&PEP		&PepsiCo Inc.						\\
				INTC		&Intel Corp.					&	&\textbf{BTI}	&\textbf{British American Tobacco}						\\
				QCOM	&QUALCOMM Inc.				&	&\textit{MO}		&\textit{Altria Group Inc.}						\\
				CRM		&Salesforce Com Inc. 			&	&GIS		&General Mills						\\
				\textbf{NVDA}		&\textbf{Nvidia Corp.}		&	&KMB		&Kimberly-Clark					\\
				\midrule
				\multicolumn{2}{l}{\textbf{Utilities} (U)} &~& \multicolumn{2}{l}{\textbf{Telecommunications Services} (TS)} \vspace{0.1cm}  \\
				Symbol	&Company					&	&Symbol		&Company						\\
				\midrule
				ETR		&Entergy Corp.					&	&T			&AT$\&$T Inc.						\\
				EXC		&Exelon Corp.					&	&VZ			&Verizon Communications			\\
				\textit{DUK}		&\textit{Duke Energy Corp. Com.} 		&	&\textit{FB}			&\textit{Facebook Inc. Com.}				\\
				FE		&FirstEnergy Corp.				&	&AMT		&American Tower Corp.				\\
				\textit{NEE}		&\textit{NextEra Energy Inc. Com.}					&	&\textbf{LUMN} 		&\textbf{Lumen Technologies}				\\
				SRE		&Sempra Energy				&	&\textbf{TMUS}			&\textbf{T-Mobile US Inc.}					\\
				\textbf{D}		&\textbf{Dominion Energy Inc.}					&	&\textbf{IQV}			&\textbf{IQVIA Holdings Inc.} 			\\
				\textbf{WEC}		&\textbf{WEC Energy Group} 		&	&\textbf{NFLX}		&\textbf{Netflix Inc.}			\\
				EIX		&Edison Int'l					&	&\textbf{BIDU}		&\textbf{Baidu Inc.}			\\
				\textit{SO}		&\textit{Southern Com.}				&	&			&									\\
				\bottomrule
			\end{tabular}
		\end{footnotesize}
	\end{center}
	\label{tab:stocks_2021}
\end{table*}


\end{appendices}

\end{document}